\documentclass[twocolumn,times,tighten]{aastex701}

\usepackage{enumitem}

\hypersetup{linkcolor=blue,citecolor=blue,filecolor=cyan,urlcolor=blue}

\def\largeplotone#1{\centering \leavevmode
\includegraphics[width=190mm]{#1} }

\def\smallplot#1{\centering \leavevmode
\includegraphics[width=85mm]{#1} }

\def\largeplotthree#1#2#3{\centering \leavevmode
\includegraphics[width=120mm]{#1}
 \hfil \includegraphics[width=120mm]{#2}
 \hfil \includegraphics[width=170mm]{#3} }
\graphicspath{{./}{figures/}}

\shorttitle{The Dusty Circumstellar Structure of the Yellow Hypergiant IRC +10420}
\shortauthors{Suh} %%
\begin{document}

\title{The Dusty Circumstellar Structure of the Yellow Hypergiant IRC +10420: Constraints from the Spectral Energy Distribution and Near-infrared Imaging Polarimetry}

\author[orcid=0000-0001-9104-9763,sname='Suh']{Kyung-Won Suh}
\affiliation{Department of Astronomy and Space Science, Chungbuk National
University, Cheongju, 28644, Republic of Korea}
\email[show]{kwsuh@chungbuk.ac.kr}

\begin{abstract}
The yellow hypergiant IRC\,+10420 is a post-red supergiant evolving toward
higher temperatures. We model its circumstellar dust with the radiative
transfer code RADMC-3D, including polarized scattering, constrained by the
spectral energy distribution (SED) from $0.4\,\mu$m to 1.3\,mm and by the $K'$
($2.2\,\mu$m) imaging polarimetry of \citet{Shenoy2015}, assuming a pole-on
view. The model consists of an equatorial structure---a flared inner disk of
$0.3\,\mu$m grains (300--3000\,au), a ring (3000--16\,000\,au) whose surface
density peaks at $\approx10^4$\,au, and a thin mid-plane disk of $300\,\mu$m
grains---embedded in a spherical envelope of $0.15\,\mu$m grains with
$\rho\propto r^{-2.3}$. It reproduces the SED within 0.07\,dex and the total-
and polarized-intensity profiles; the fractional polarization peaks at 15\% at
$1\farcs7$ (observed: 14\%). The rising surface density of the ring produces
the polarized-intensity plateau, and $0.5\,\mu$m grains at its outer edge the
decline beyond $2''$; grains of $\gtrsim0.7\,\mu$m are excluded. A thin disk of
large grains suggests a settled mid-plane. Stellar temperatures of
7000--8500\,K fit equally well. Of the total extinction ($A_V\approx6.9$\,mag),
5\,mag is interstellar. Most of the $7.4\times10^{-2}\,M_\odot$ of dust is
equatorial; the envelope implies a mass-loss rate of
$\approx10^{-5}$--$10^{-4}\,(\psi/100)\,M_\odot\,\mathrm{yr^{-1}}$. Because
equatorial dust redirects starlight poleward, the pole-on SED overestimates the
luminosity by $\approx1.5$; at the \textit{Gaia} distance,
$L_\star\approx3.3\times10^5\,L_\odot$. Subtracting a scaled stellar PSF
removes most of the nebular light within $\approx1''$.
\end{abstract}



\section{Introduction}\label{sec:intro}

Massive stars with initial masses of roughly $20$--$40\,M_\odot$ lose a large
fraction of their hydrogen envelopes during and after the red supergiant (RSG)
phase. How much mass they lose, and whether it is lost steadily or in brief
episodes, largely determines their subsequent evolution and the type of supernova
they produce \citep[e.g.,][]{Smith2014}. Some of these stars evolve back toward
higher effective temperatures after the RSG phase. While crossing the
intermediate-temperature region of the Hertzsprung--Russell diagram, they appear
as yellow hypergiants (YHGs). In this region, the ``yellow void'' of
\citet{deJager1998}, the stellar atmosphere is thought to become dynamically
unstable, and the mass loss is expected to increase sharply
\citep{deJager1998,Humphreys1994}. The circumstellar envelopes (CSEs) of YHGs
therefore record the late mass-loss history of massive stars, which current
evolutionary models describe only poorly.

IRC\,+10420 (V1302~Aql) is one of the most extreme and best-studied members of
this class. It is a bright near-infrared source and one of the warmest stars
known to host OH masers \citep{Giguere1976}. Its spectral type has changed from
F8\,Ia$^{+}$ in the early 1970s \citep{Humphreys1973} to an early A type. This
implies an increase in effective temperature within a few decades
\citep{Oudmaijer1996,Oudmaijer1998,Klochkova1997,Klochkova2002,Klochkova2016};
since the mid-1990s, however, the spectral type has remained nearly constant
\citep{Koumpia2022}. IRC\,+10420 is thus widely regarded as a post-RSG caught in
the act of crossing the yellow void. Earlier distance estimates ranged from about
3 to 6\,kpc \citep{Jones1993,Oudmaijer1996,Humphreys2002}. The \textit{Gaia}
parallax gives $d\simeq4.3$\,kpc and $L\approx5\times10^{5}\,L_\odot$
\citep{Koumpia2022}. The star therefore lies close to the empirical upper
luminosity limit for cool supergiants \citep{Humphreys1979,Humphreys1994} if its
emission is isotropic (see Section~\ref{sec:luminosity}).

IRC\,+10420 is surrounded by a massive, optically thick dusty envelope and is one
of the brightest mid-infrared sources in the sky. Its CSE has been observed on
scales from milliarcseconds to tens of arcseconds. Imaging with the
\textit{Hubble Space Telescope} (HST) revealed a complex nebula within $\sim2''$
of the star. The nebula contains numerous condensations arranged in jet-like
structures, rays, arcs, and partial shells. HST imaging also revealed fainter,
roughly spherical outer shells at $5''$--$6''$ \citep{Humphreys1997,Tiffany2010}.
These structures point to several episodes of high mass loss over the past few
centuries to millennia \citep{Humphreys1997,Humphreys2002,Tiffany2010}.
Interferometric CO maps show a detached molecular envelope with an inner cavity,
which indicates a recent drop in the mass-loss rate
\citep{CastroCarrizo2007,DinhVTrung2009}. Mid- and far-infrared imaging suggests
the same history. The average mass-loss rate was
$\sim2\times10^{-3}\,M_\odot\,\mathrm{yr}^{-1}$ until $\sim2000$\,yr ago and then
decreased by about an order of magnitude \citep{Shenoy2016}. Near-infrared
speckle interferometry combined with spherically symmetric radiative transfer
(RT) modelling likewise points to a detached dust shell. It also indicates a
superwind phase that ended in the recent past \citep{Blocker1999}.

The three-dimensional geometry of the ejecta is still debated. \citet{Jones1993}
proposed a rotating equatorial disk viewed at an intermediate inclination. In
contrast, HST long-slit spectroscopy and integral-field spectroscopy favour an
axisymmetric outflow, either bipolar or equatorially enhanced, seen nearly
pole-on \citep{Humphreys2002,Davies2007}. The proper motions of the inner ejecta
measured from second-epoch HST imaging support the same picture
\citep{Tiffany2010}. Near-infrared spectro-interferometry of the
Br$\gamma$-emitting wind is also consistent with a compact bipolar outflow viewed
close to its axis \citep{deWit2008,Oudmaijer2013,Koumpia2022}. Long-term optical
(spectro)polarimetry shows variable intrinsic polarization. This variability has
been attributed to an axisymmetric structure oriented along the short axis of the
reflection nebula \citep{Patel2008}.

Scattered-light polarimetry is a particularly direct probe of this geometry.
Polarimetric, coronagraphic near-infrared imaging first revealed an extended,
nearly circular reflection nebula around IRC\,+10420
\citep{KastnerWeintraub1995}. More recently, adaptive-optics imaging polarimetry
at $2.2\,\mu$m showed a centrosymmetric polarization pattern. After subtraction
of the stellar point-spread function (PSF), the fractional polarization of the
nebular light reaches $\sim30\%$ at $1''$--$2''$ from the star
\citep{Shenoy2015}. Such strong polarization requires scattering angles near
$90^\circ$. The scattering dust must therefore lie mostly close to the plane of
the sky, as expected for a system viewed pole-on. The polarization also depends
on the size distribution of the grains, which is poorly constrained by the
thermal emission alone.

RT models of IRC\,+10420 have so far been constrained mainly by the spectral
energy distribution (SED). In some cases, they have also been constrained by
high-resolution near-infrared intensity profiles
\citep[e.g.,][]{Oudmaijer1996,Blocker1999}. The SED of an optically thick dusty
envelope, however, is degenerate in geometry, optical depth, and grain
properties. Spherical and strongly non-spherical configurations can often
reproduce the same SED equally well. In contrast, maps of the scattered-light
intensity and polarization respond very differently to the spatial distribution
of the dust, to the viewing angle, and to the grain size. Modelling the SED
together with resolved imaging polarimetry therefore offers a way to break these
degeneracies. To our knowledge, no such joint analysis of IRC\,+10420 has been
published.

In this paper, we model the dusty CSE of IRC\,+10420 with the three-dimensional
Monte Carlo RT code RADMC-3D \citep{Dullemond2012}, including full polarized
scattering. The models are constrained simultaneously by the SED from $0.4\,\mu$m
to 1.3\,mm and by the $K'$ ($2.2\,\mu$m) adaptive-optics imaging polarimetry of
\citet{Shenoy2015}. We combine a spherical envelope with flared, disk-like
density components and determine (1) the radial distribution and mass of the
dust, (2) the grain size in each component, (3) the division of the extinction
between the interstellar medium and the circumstellar envelope, and (4) the
effect of the dust geometry on the luminosity inferred from the SED. From these
results, we estimate the mass-loss rate of the expanding envelope and constrain
the origin of the equatorial structure.

The paper is organized as follows. Section~\ref{sec:obs} describes the
observational data. Section~\ref{sec:model} presents the RT model.
Section~\ref{sec:results} compares the model with the observed SED and with the
polarimetric profiles and images. Section~\ref{sec:discussion} discusses the
mass-loss history and the geometry of the ejecta. Section~\ref{sec:summary}
summarizes our conclusions.

\section{Observational Data}
\label{sec:obs}

\begin{deluxetable*}{rllcccl}
\tabletypesize{\footnotesize} \tablecaption{Observational Data of IRC\,+10420
Used in the SED Fit \label{tab:obs}} \tablehead{ \colhead{$\lambda$} &
\colhead{Band} & \colhead{Instrument} & \colhead{$\lambda F_\lambda$} &
\colhead{$\sigma(\lambda F_\lambda)$} &
\colhead{Epoch} & \colhead{Reference} \\
\colhead{($\mu$m)} & \colhead{} & \colhead{} & \colhead{(W~m$^{-2}$)} &
\colhead{(W~m$^{-2}$)} & \colhead{} & \colhead{} } \startdata
\multicolumn{7}{c}{Photometry of the 1990s and far-infrared/submillimeter data} \\
\hline
0.43 & $B$ & ground-based & $7.31\times10^{-14}$ & $7.3\times10^{-15}$ & 1990s & \citet{Oudmaijer1996} \\
0.55 & $V$ & ground-based & $8.52\times10^{-13}$ & $4.3\times10^{-14}$ & 1990s & \citet{Jones1993,Oudmaijer1996}\tablenotemark{a} \\
0.65 & $R$ & ground-based & $2.20\times10^{-12}$ & $1.1\times10^{-13}$ & 1990s & \citet{Jones1993} \\
0.77 & $I$ & ground-based & $6.24\times10^{-12}$ & $3.1\times10^{-13}$ & 1990s & \citet{Jones1993} \\
1.26 & $J$ & ground-based & $2.32\times10^{-11}$ & $1.2\times10^{-12}$ & 1990s & \citet{Oudmaijer1996} \\
1.65 & $H$ & ground-based & $2.94\times10^{-11}$ & $1.5\times10^{-12}$ & 1990s & \citet{Oudmaijer1996} \\
2.21 & $K$ & ground-based & $3.67\times10^{-11}$ & $1.8\times10^{-12}$ & 1990s & \citet{Oudmaijer1996} \\
3.83 & $L$ & ground-based & $7.43\times10^{-11}$ & $5.2\times10^{-12}$ & 1990s & \citet{Oudmaijer1996} \\
12 & 12~$\mu$m & IRAS & $3.36\times10^{-10}$ & $3.4\times10^{-11}$ & 1983 & \citet{Beichman1988} \\
12.1 & $C$ & MSX & $3.15\times10^{-10}$ & $2.2\times10^{-11}$ & 1996--1997 & \citet{Egan2003} \\
14.7 & $D$ & MSX & $2.65\times10^{-10}$ & $1.9\times10^{-11}$ & 1996--1997 & \citet{Egan2003} \\
21.3 & $E$ & MSX & $3.19\times10^{-10}$ & $2.2\times10^{-11}$ & 1996--1997 & \citet{Egan2003} \\
25 & 25~$\mu$m & IRAS & $2.78\times10^{-10}$ & $2.8\times10^{-11}$ & 1983 & \citet{Beichman1988} \\
60 & 60~$\mu$m & IRAS & $3.60\times10^{-11}$ & $5.4\times10^{-12}$ & 1983 & \citet{Beichman1988} \\
100 & 100~$\mu$m & IRAS & $6.00\times10^{-12}$ & $1.2\times10^{-12}$ & 1983 & \citet{Beichman1988} \\
65 & \textit{N60} & AKARI/FIS & $2.29\times10^{-11}$ & $3.4\times10^{-12}$ & 2006--2007 & \citet{Yamamura2010} \\ %% CHECK: AKARI values taken from the previous obs.dat
90 & \textit{WIDE-S} & AKARI/FIS & $1.41\times10^{-11}$ & $2.1\times10^{-12}$ & 2006--2007 & \citet{Yamamura2010} \\
140 & \textit{WIDE-L} & AKARI/FIS & $1.43\times10^{-12}$ & $4.3\times10^{-13}$ & 2006--2007 & \citet{Yamamura2010} \\
160 & \textit{N160} & AKARI/FIS & $9.36\times10^{-13}$ & $2.8\times10^{-13}$ & 2006--2007 & \citet{Yamamura2010} \\
450 & 450~$\mu$m & JCMT/SCUBA & $7.55\times10^{-15}$ & $1.5\times10^{-15}$ & 1997--2005 & \citet{DiFrancesco2008} \\
870 & 850~$\mu$m & JCMT/SCUBA & $8.37\times10^{-16}$ & $1.7\times10^{-16}$ & 1997--2005 & \citet{DiFrancesco2008} \\
1300 & 1.3~mm & IRAM 30~m & $2.33\times10^{-16}$ & $2.1\times10^{-17}$ & 1990 & \citet{Walmsley1991} \\ %% CHECK: telescope (IRAM 30 m?) and epoch
\hline
\multicolumn{7}{c}{Photometry of the 2000s} \\
\hline
0.4442 & $B$ & APASS & $1.33\times10^{-13}$ & $4.8\times10^{-15}$ & 2009--2014 & \citet{Henden2016} \\
0.5537 & $V$ & APASS & $8.55\times10^{-13}$ & $1.1\times10^{-14}$ & 2009--2014 & \citet{Henden2016} \\
1.25 & $J$ & 2MASS & $2.52\times10^{-11}$ & $7.3\times10^{-13}$ & 1998--2000 & \citet{Skrutskie2006} \\
1.63 & $H$ & 2MASS & $2.91\times10^{-11}$ & $1.7\times10^{-12}$ & 1998--2000 & \citet{Skrutskie2006} \\
\hline
\multicolumn{7}{c}{Spectroscopy} \\
\hline
2.36--45.4 & \nodata & ISO/SWS & \nodata & \nodata & 1996--1997 & \citet{Sloan2003} \\ %% CHECK: reference / TDT of the SWS spectrum
\enddata
\tablecomments{The fluxes are apparent values, i.e., not corrected for
interstellar extinction; in the fit they are corrected for $A_V=5$~mag with
$R_V=3.1$ (Section~\ref{sec:extinction}). The photometric points and the ISO/SWS
spectrum receive equal total weight in the fit (Section~\ref{sec:sed}). Saturated
measurements (2MASS $K_s$ and WISE $W1$--$W4$) and the Gaia $G$, $G_{\rm BP}$,
and $G_{\rm RP}$ bands, whose effective wavelengths are poorly defined for this
strongly reddened star, are not used.} \tablenotetext{a}{Mean of
\citet{Jones1993} and \citet{Oudmaijer1996}.}
\end{deluxetable*}

\subsection{Spectral energy distribution} \label{sec:obs_sed}

The photometric and spectroscopic data used in the SED fit are listed in
Table~\ref{tab:obs}. They are taken from the literature and public catalogs. The
core of the data set is the ISO/SWS spectrum \citep[2.36--45.4\,$\mu$m,
1996--1997;][]{Sloan2003}. We combine it with ground-based $BVRIJHKL$ photometry
of the 1990s \citep{Jones1993,Oudmaijer1996}, which is roughly contemporaneous
with the spectrum. We add mid- to far-infrared photometry from IRAS
\citep{Beichman1988}, MSX \citep{Egan2003}, and AKARI/FIS \citep{Yamamura2010}.
At longer wavelengths we use submillimeter fluxes from JCMT/SCUBA
\citep{DiFrancesco2008} and the 1.3\,mm flux measured with the IRAM 30\,m
telescope \citep{Walmsley1991}.

To check for long-term changes, we also use unsaturated photometry of the 2000s:
APASS $B$ and $V$ \citep{Henden2016} and 2MASS $J$ and $H$ \citep{Skrutskie2006}.
These agree with the 1990s photometry within 0.11\,dex (Section~\ref{sec:sed}).
We do not use the saturated 2MASS $K_s$ and WISE W1--W4 measurements. We also
exclude the Gaia $G$, $G_{\rm BP}$, and $G_{\rm RP}$ bands, whose effective
wavelengths are poorly defined for this strongly reddened star.

The data span more than three decades, from IRAS in 1983 to AKARI in 2006--2007.
The dust emission at $\lambda\gtrsim10\,\mu$m arises at $r\gtrsim300$\,au. At an
expansion velocity of $\sim35$\,km\,s$^{-1}$, the dust moves only $\sim70$\,au
per decade, which is small compared with the scales probed by the model. We
therefore fit all data with a single model. The optical variability of the star,
$\lesssim0.2$--0.3\,mag in the $B$ band \citep[e.g.,][]{Patel2008}, is comparable
to the adopted flux uncertainties, and the optical brightness has remained nearly
constant since the 1970s (Appendix~\ref{sec:ltphot}).

The far-infrared data were obtained with large beams or detector apertures:
$\approx0.75'\times4.5'$ to $3'\times5'$ for IRAS \citep{Beichman1988} and
37$''$--61$''$ (FWHM) for AKARI/FIS \citep{Kawada2007}. The ISO/SWS apertures
($14''\times20''$ to $20''\times33''$) are also larger than the nebula. All of
these encompass the entire model nebula. The SCUBA and IRAM beams are smaller:
7\farcs5 at 450\,$\mu$m, 14$''$ at 850\,$\mu$m, and $\approx11''$ at 1.3\,mm. In
the model, however, 98\%--99\% of the emission at these wavelengths arises within
the beam FWHM of the star. Weighting the model images with Gaussian beams of
these FWHMs reduces the fluxes by only 9\%, 3\%, and 4\% at 450\,$\mu$m,
850\,$\mu$m, and 1.3\,mm, respectively. These reductions correspond to
$\leq0.04$\,dex, smaller than the observational uncertainties. We therefore
compare all observed fluxes with the total model flux.

\subsection{Near-infrared imaging polarimetry} \label{sec:obs_pol}

For the scattered light, we use the $K'$ ($\lambda_0=2.20\,\mu$m,
$\Delta\lambda=0.12\,\mu$m) adaptive-optics imaging polarimetry of
\citet{Shenoy2015}, obtained with MMT-Pol on 2012 September 26 (UT) with a pixel
scale of $0\farcs043$. The central point source is saturated at
$r\lesssim0\farcs4$, and the stellar PSF used in the reduction is scaled at
$0\farcs5$; we therefore compare the profiles only at $r\geq0\farcs7$.

Following \citet{Shenoy2015}, we use four radial profiles averaged in $0\farcs2$
annuli within the east and west sectors: the total intensity $I^{\rm raw}$, the
polarized intensity $I_P=(Q^2+U^2)^{1/2}$, the fractional polarization $p^{\rm
raw}=I_P/I^{\rm raw}$, and the ratio $I_P/I^\star$, where $I^\star$ is the
stellar PSF scaled to $I^{\rm raw}$ at $0\farcs5$. The profiles were digitized
from Figure~6 of \citet{Shenoy2015}. Because the absolute flux calibration is not
used, all intensities are normalized to $I^{\rm raw}$ at $0\farcs7$. The
convolution of the model images with the PSF and their reduction in the same way
as the observations are described in Section~\ref{sec:images}.

\subsection{Distance} \label{sec:obs_dist}

We adopt $d=5$\,kpc, the distance used in most previous studies of the
circumstellar envelope
\citep[e.g.,][]{Jones1993,Blocker1999,CastroCarrizo2007,DinhVTrung2009}, which
allows direct comparison of physical scales and mass-loss rates. The
\textit{Gaia} parallax gives $d=4.26^{+0.88}_{-0.75}$\,kpc
\citep{BailerJones2021,Koumpia2022}, consistent with 5\,kpc within its
uncertainty. For a different distance and fixed angular scales, the luminosity
and the dust masses scale as $d^2$, and the linear scales, the mass-loss rates,
and the kinematic ages (for a fixed expansion velocity) scale as $d$. At
$d=4.26$\,kpc, the adopted model corresponds to
$L_\star\approx3.3\times10^5\,L_\odot$.

\section{Radiative transfer model} \label{sec:model}

We model the spectral energy distribution (SED) and the $K'$ imaging polarimetry
of IRC\,+10420 with the three-dimensional Monte Carlo radiative transfer code
RADMC-3D \citep{Dullemond2012}, including full polarized scattering with the
Mueller matrices of spherical grains. The dust temperature is computed
self-consistently from the stellar radiation field, separately for each grain
population, using $2\times10^{6}$ photon packages and the modified random walk
method in optically thick regions. The spectra are computed with $10^{6}$
scattering photon packages at each of the 91 wavelengths of the model grid.

The spherical grid has $150\times180\times60$ cells in $(r,\theta,\phi)$,
logarithmically spaced in $r$ between 300 and $5\times10^5$\,au and uniformly
spaced in $\theta$ (1$^\circ$). The flared disks are well resolved; the thin disk
of large grains ($H/r\approx0.007$--0.02, i.e., 0.4$^\circ$--1.2$^\circ$)
occupies only the one to three cells closest to the equatorial plane, so its
vertical temperature structure is not resolved. A test with 0.25$^\circ$ cells
within $5^\circ$ of the equatorial plane gives the same SED and changes the rms
residuals of the $K'$ profiles by $\leq0.003$\,dex. Its dust mass, computed from
the gridded density, is that listed in Table~\ref{tab:density}. We adopt a
distance of $d=5$~kpc (Section~\ref{sec:obs_dist}) and view the system pole-on
($i=0^\circ$).

\subsection{Modeling approach} \label{sec:approach}

The adopted model was not obtained by a formal minimization: with five
components, each described by a radial law, radial limits, a scale height, and a
grain population, the parameter space is too large. Instead, the model was built
step by step, each component being introduced to reproduce a specific observable,
and its parameters were adjusted with small grids of models around the current
best model. (1) A spherical envelope that produces the observed infrared excess
absorbs most of the starlight in all directions and therefore extinguishes the
star far more than the observed circumstellar reddening allows; the dust must be
concentrated toward the equatorial plane, which we view pole-on. (2) A flared
inner disk provides the warm dust ($\lesssim3000$\,au) required by the near- and
mid-infrared SED. (3) The nearly flat polarized-intensity profile between
$0\farcs7$ and $1\farcs7$ requires a surface density of the scattering grains
that rises outward, which we describe as a detached ring; the decline beyond
$2''$ requires a steep outer edge of weakly polarizing grains
(Section~\ref{sec:images}). (4) A geometrically thin disk of large grains is
needed for the (sub)millimeter fluxes, which submicron grains at the temperatures
of the model underpredict. (5) Finally, the stellar temperature, the interstellar
extinction, and the envelope grains were determined together from the optical
colors (Sections~\ref{sec:extinction} and \ref{sec:sed}).

The resulting model reproduces the data but is not unique. Other geometries,
e.g., the walls of a bipolar cavity instead of a flared ring
(Section~\ref{sec:geometry}), or other combinations of density laws and grain
sizes may fit equally well. The tests in Tables~\ref{tab:tstar} and
\ref{tab:ringtest} show which properties are constrained: the pole-on,
equatorially concentrated geometry, the rising surface density of the ring and
its steep outer edge, the grain sizes in each component, and the luminosity.

As an example, we replaced the ring and its outer edge by the walls of a bipolar
cavity: bands of $\pm15^\circ$ in latitude centered at a latitude $b$ above the
equatorial plane, with the same grains, radial mass distributions, and dust
masses as the ring and its outer edge. This geometry resembles a typical
hourglass more closely than the adopted model does. Seen pole-on, a wall at
latitude $b$ scatters starlight toward the observer through $90^\circ-b$, so
walls at high latitudes scatter forward and polarize weakly. For $b=20^\circ$,
the SED is fitted nearly as well ($\sigma_{\log}=0.076$\,dex), but the polarized
intensity is too high within $1''$ and too low beyond $2''$, and the rms
residuals of the $K'$ profiles increase to 0.14--0.31\,dex. For $b=35^\circ$ and
$50^\circ$, both the SED ($\sigma_{\log}=0.093$ and 0.087\,dex) and the
polarimetry are fitted worse; at $b=50^\circ$, $p^{\rm raw}\leq0.04$ at all
radii. At $b=65^\circ$, the walls also obscure the star ($A_V^{\rm cs}=2.8$\,mag,
$\sigma_{\log}=0.147$\,dex), and $p^{\rm raw}\leq0.01$. The polarimetry thus
requires the scattering surface to lie close to the equatorial plane
($\lesssim20^\circ$--$30^\circ$), which excludes a narrow bipolar cavity. A
wide-angle cavity with walls at such low latitudes is essentially the geometry of
the adopted model, in which the flared surface of the ring rises to
$20^\circ$--$30^\circ$ above the equatorial plane and bounds biconical,
low-density regions around the poles (Section~\ref{sec:geometry}). The flared
ring and a wide-angle cavity are thus two descriptions of nearly the same
structure.

\subsection{Central star} \label{sec:star}

The spectroscopic effective temperature of IRC\,+10420 increased between the
1970s and the 1990s. The star was classified as F8\,Ia$^+$ in the early 1970s
\citep{Humphreys1973}. Spectra from the 1990s, contemporaneous with most of our
SED data, give an A-type spectrum and $T_{\rm eff}\approx7900$--8500\,K
\citep{Klochkova1997,Nieuwenhuijzen2000}. Although later spectra were interpreted
as indicating a further increase \citep{Klochkova2002,Klochkova2016},
\citet{Koumpia2022} found that the spectral type, and thus the temperature, has
remained essentially constant between 1994 and 2019. The SED data, obtained
mostly between 1983 and 2007, therefore correspond to a single stellar
temperature.

The spectrum, however, forms in a dense wind. \citet{Humphreys2002} argued that
the wind is optically thick. If so, the observed spectral type and its changes
reflect the conditions in the wind rather than those of the hydrostatic stellar
surface. For the dust radiative transfer, the relevant quantity is the spectrum
of the radiation that emerges from the star and its ionized wind. The wind lies
well inside the inner radius of the dust (300\,au), so the star and wind can be
treated as a point source.

We represent this source by a blackbody with $T_\star$ and $L_\star$ as free
parameters. The dust emission depends mainly on $L_\star$. It is nearly
independent of $T_\star$, because the dust reprocesses the absorbed starlight
into the infrared regardless of its original spectrum. The optical and
near-infrared SED, in contrast, depends on $T_\star$, but this dependence is
degenerate with the reddening. A hotter star is intrinsically bluer. To match the
same observed colors, it therefore requires either more interstellar extinction
or more selective circumstellar extinction. We therefore determine $T_\star$
together with the interstellar extinction and the envelope grains
(Section~\ref{sec:extinction}).

The adopted model has $T_\star=7000$\,K. Its luminosity,
$L_\star=4.5\times10^5\,L_\odot$, is fixed by the scale factor $s=1.01$; this and
the models with $T_\star=8000$ and 8500\,K are discussed in
Section~\ref{sec:sed}.

A blackbody reproduces neither the Balmer jump nor the line blanketing of an
A-type photosphere. We therefore also tested the model atmospheres of
\citet{Kurucz1992} with solar metallicity, $T_{\rm eff}=7000$\,K and $\log
g=0.5$, and $T_{\rm eff}=8000$\,K and $\log g=1.0$ (the lowest gravity available
at this temperature), normalized to the same luminosity and with all other model
parameters unchanged. They give clearly poorer fits than blackbodies of the same
temperature: $\sigma_{\log}=0.083$ and 0.085\,dex ($\chi^2_\nu=1.41$ and 1.40),
compared with 0.068\,dex ($\chi^2_\nu=1.20$) for both blackbodies. The model
atmospheres are too bright in the $B$ band by 0.15--0.17\,dex, i.e., their
optical colors are bluer than observed, while the luminosity scale is nearly
unchanged ($s=1.00$--1.02). This is consistent with the suggestion that the
spectrum of IRC\,+10420 forms in an optically thick wind \citep{Humphreys2002},
whose emergent continuum need not show the features of a hydrostatic photosphere.
Plane-parallel, hydrostatic model atmospheres are in any case of limited validity
for such an extended, wind-dominated atmosphere. We therefore represent the star
by a blackbody.

\begin{figure}
\smallplot{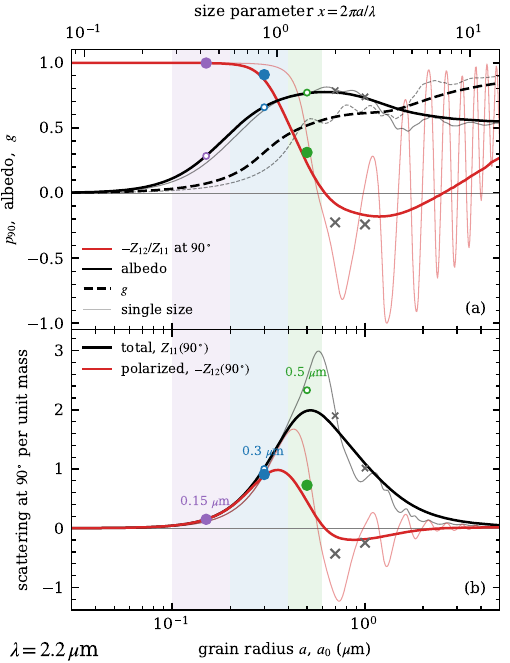}
\caption{Scattering properties of spherical silicate grains \citep{Suh1999}
at $\lambda=2.2\,\mu$m as a function of grain radius (bottom axis) and size
parameter $x=2\pi a/\lambda$ (top axis).
(a) Linear polarization of light scattered through $90^\circ$,
$-Z_{12}/Z_{11}$ (negative values correspond to polarization parallel to
the scattering plane, i.e., radial polarization vectors), albedo, and
asymmetry parameter $g$; (b) total ($Z_{11}$) and polarized ($-Z_{12}$) intensity
scattered through $90^\circ$ per unit dust mass, normalized to the
$0.3\,\mu$m distribution. Thick lines are for Gaussian size distributions
with $\sigma_a=a_0/3$, plotted against $a_0$; thin lines are for single
grain sizes and show the Mie resonances at $x\gtrsim2$. Symbols show the
distributions of Section~\ref{sec:dust} (circles: used in the model, with shaded bands
marking $a_0\pm\sigma_a$; crosses: the rejected 0.7 and $1\,\mu$m
distributions). The 0.5--1\,$\mu$m distributions are narrower than
$\sigma_a=a_0/3$, so their symbols lie off the thick lines.
\label{fig:polsize}}
\end{figure}

\subsection{Dust properties} \label{sec:dust}

All grains are spherical silicate grains with the optical constants of cold
circumstellar silicate \citep{Suh1999} and a material density of 3.3~g~cm$^{-3}$.
Absorption and scattering cross sections and the scattering matrices are computed
with Mie theory \citep{Bohren1983} and averaged over a Gaussian distribution of
grain radii,
\begin{equation}
  w(a)\propto\exp\left[-\frac{(a-a_0)^2}{2\sigma_a^2}\right],
  \qquad a_{\rm min}\le a\le a_{\rm max},
  \label{eq:gauss}
\end{equation}
where $w(a)\,da$ is the fraction of the dust mass in grains with radii between
$a$ and $a+da$. The opacities per unit dust mass are averaged with the weights
$w(a)$, and the asymmetry parameter $g=\langle\cos\theta\rangle$ is weighted by
the scattering cross section, so that $g$ is consistent with the scattering
matrix.

We use four grain populations (Figure~\ref{fig:polsize}). The envelope consists
of small grains with $a_0=0.15~\mu$m and $\sigma_a=0.05~\mu$m ($0.01\le
a\le0.29~\mu$m). The disk components consist of larger grains: $a_0=0.3~\mu$m and
$\sigma_a=0.1~\mu$m ($0.1\le a\le0.5~\mu$m) in the inner disk and the ring;
$a_0=0.5~\mu$m and $\sigma_a=0.1~\mu$m ($0.2\le a\le0.8~\mu$m) in the outer edge
of the ring; and $a_0=300~\mu$m and $\sigma_a=50~\mu$m ($200\le a\le400~\mu$m) in
the thin disk. These sizes are consistent with grain growth in the dense
equatorial material, which is quasi-stationary or flows outward more slowly than
the envelope (see Section~\ref{sec:massloss}).

Because the Gaussian describes the mass distribution, the corresponding number
distribution ($\propto w(a)/a^3$) is skewed toward smaller grains; the
area-weighted radii $a_{\rm eff}=\langle a^3\rangle/\langle a^2\rangle$, which
determine the optical properties, are 0.13, 0.27, 0.48, and 290~$\mu$m for the
0.15, 0.3, 0.5, and 300~$\mu$m populations.

For the size distributions of the model, the optical extinction of the
0.15~$\mu$m grains has a wavelength dependence similar to that of interstellar
dust ($\kappa^{\rm ext}_B/\kappa^{\rm ext}_V=1.33$, equivalent to $R_V=3.0$).
Their albedo at $K'$ is low (0.28), so they contribute little to the $K'$ nebula.
The 0.3 and 0.5~$\mu$m grains have high albedos (0.66 and 0.77) and dominate the
scattered light; they differ mainly in its polarization. The size parameter
$x=2\pi a/\lambda$ at $K'$ is $\approx0.9$ for the 0.3~$\mu$m grains, and light
scattered through $90^\circ$ is polarized by 91\%, close to the Rayleigh limit.
For the 0.5~$\mu$m grains ($x\approx1.4$), the polarization at $90^\circ$ drops
to 31\%, while the scattering opacity per unit dust mass is 2.8 times larger and
the scattering is more forward-directed ($g=0.52$ instead of 0.29). Per unit dust
mass, the 0.5~$\mu$m grains therefore scatter 2.3 times more light through
$90^\circ$ than the 0.3~$\mu$m grains, but produce only 0.8 times as much
polarized light (Figure~\ref{fig:polsize}b). For $x\gtrsim2$
($a\gtrsim0.7\,\mu$m), the polarization at $90^\circ$ becomes negative or close
to zero (Section~\ref{sec:images}).

The 300~$\mu$m grains are in the geometric-optics limit at all wavelengths
shorter than $\approx100~\mu$m. Their diffraction peak, of angular width
$\sim\lambda/(2\pi a)$, is much narrower than the $1^\circ$ angular grid of the
scattering matrix. We therefore truncate the scattering matrix within $3^\circ$
of the forward direction and treat the light in the removed peak as unscattered,
which reduces the scattering opacity and the asymmetry parameter but conserves
the transport opacity $\kappa^{\rm sca}(1-g)$. These grains are confined to the
thin equatorial disk and contribute little to the scattered light.

We use Gaussian rather than power-law size distributions. Power laws such as that
of \citet{Mathis1977} describe interstellar dust that has been processed by
shattering and coagulation, whereas freshly formed circumstellar dust grows from
nucleation in the outflow and is expected to have a narrower size distribution,
as found in dynamical wind models \citep{Hofner2008} and in scattered-light
polarimetry of evolved stars \citep{Norris2012,Scicluna2015}. A Gaussian also
characterizes each component by a single representative grain size; for a power
law, the scattered light at $K'$ would be dominated by grains near the upper
cutoff, which would then play the role of $a_0$.

The conclusions do not depend critically on the width of the distributions. For
distributions truncated at $\pm3\sigma_a$, increasing $\sigma_a$ from $0.1\,\mu$m
to $a_0/2$ lowers the polarization at $90^\circ$ at $K'$ from 0.87 to 0.69 for
$a_0=0.3\,\mu$m and from 0.31 to 0.17 for $a_0=0.5\,\mu$m, so that the larger
grains at the outer edge of the ring remain the more weakly polarizing ones. For
$a_0=0.7$ and $1\,\mu$m it remains between $-0.37$ and 0
(Figure~\ref{fig:polsize}).

\begin{deluxetable*}{lllcccc}
\tabletypesize{\footnotesize} \tablecaption{Dust Components of the Adopted Model
\label{tab:density}} \tablehead{ \colhead{Component} & \colhead{Grains} &
\colhead{$R_{\rm in}$--$R_{\rm out}$} & \colhead{Radial Law} &
\colhead{Normalization} & \colhead{$h_0$} &
\colhead{$M_{\rm dust}$} \\
\colhead{} & \colhead{} & \colhead{(au)} & \colhead{} & \colhead{} & \colhead{} &
\colhead{($M_\odot$)} } \startdata
1 Inner disk        & 0.3~$\mu$m  & 300--3000            & $\Sigma\propto r^{-1}$   & $\Sigma_{1,0}=7.6\times10^{-3}$~g~cm$^{-2}$           & 0.10    & $1.49\times10^{-2}$ \\
2 Large-grain disk  & 300~$\mu$m  & 300--12,000         & $\Sigma\propto r^{-1}$   & $\Sigma_{2,0}=1.64\times10^{-3}$~g~cm$^{-2}$          & 0.01    & $1.38\times10^{-2}$ \\
3 Detached ring     & 0.3~$\mu$m  & 3000--10,000        & $\Sigma\propto r^{+1.5}$ & $\Sigma_{b}=9.90\times10^{-4}$~g~cm$^{-2}$            & 0.10    & $2.00\times10^{-2}$ \\
4 Ring outer edge   & 0.5~$\mu$m  & 10,000--16,000     & $\Sigma\propto r^{-6}$   & $\Sigma_{b}=9.90\times10^{-4}$~g~cm$^{-2}$            & 0.10    & $1.42\times10^{-2}$ \\
5 Envelope          & 0.15~$\mu$m & 500--500,000        & $\rho\propto r^{-2.3}$   & $\rho_{{\rm env},0}=2.34\times10^{-20}$~g~cm$^{-3}$   & \nodata & $1.10\times10^{-2}$ \\
\hline
Total               &             &                      &                          &                                                       &         & $7.4\times10^{-2}$  \\
\enddata
\tablecomments{$\Sigma_{k,0}$ is given at $R_0=1000$~au, $\Sigma_b$ at
$R_b=10\,000$~au, and $\rho_{{\rm env},0}$ at $R_{\rm env}=500$~au
(Equations~\ref{eq:s1}--\ref{eq:env}). The flaring index is $\alpha=2/7$ for all
disk components (Equation~\ref{eq:scaleheight}). Grain sizes are the centers of
the Gaussian size distributions (Section~\ref{sec:dust}). Central star:
$L_\star=4.5\times10^{5}\,L_\odot$, $T_\star=7000$~K; $d=5$~kpc, $i=0^\circ$;
interstellar extinction $A_V=5.0$~mag \citep{Blocker1999,DinhVTrung2009},
$R_V=3.1$.}
\end{deluxetable*}

\subsection{Dust density distribution} \label{sec:density}

The circumstellar dust is described by five components: four disk-like components
that share a common vertical structure but follow different radial laws, and a
spherical envelope. We use spherical coordinates $(r,\theta,\phi)$ with the polar
axis along the line of sight, so that $z=r\cos\theta$ is the height above the
equatorial plane.

\paragraph{Disk components}
Each disk component $k$ ($k=1,\ldots,4$) is a flared structure with a Gaussian
vertical profile \citep{Kenyon1987},
\begin{equation}
  \rho_k(r,\theta) = \frac{\Sigma_k(r)}{\sqrt{2\pi}\,H_k(r)}
  \exp\left[-\frac{1}{2}\left(\frac{r\cos\theta}{H_k(r)}\right)^{2}\right]
  \label{eq:disk}
\end{equation}
for $R_{{\rm in},k}\le r\le R_{{\rm out},k}$, and $\rho_k=0$ outside this range.
The scale height is
\begin{equation}
  \frac{H_k(r)}{r} = h_{0,k}\left(\frac{r}{R_0}\right)^{\alpha},
  \label{eq:scaleheight}
\end{equation}
with $R_0=1000$~au and $\alpha=2/7$. The four components differ in their
surface-density laws and dust populations:
\begin{eqnarray}
  \Sigma_1(r) &=& \Sigma_{1,0}\,(r/R_0)^{-1}, \label{eq:s1}\\
  \Sigma_2(r) &=& \Sigma_{2,0}\,(r/R_0)^{-1}, \label{eq:s2}\\
  \Sigma_3(r) &=& \Sigma_{b}\,(r/R_b)^{+1.5}, \label{eq:s3}\\
  \Sigma_4(r) &=& \Sigma_{b}\,(r/R_b)^{-6},   \label{eq:s4}
\end{eqnarray}
defined over $300$--$3000$, $300$--$12\,000$, $3000$--$10\,000$, and
$10\,000$--$16\,000$~au, respectively. Here $R_b=10\,000$~au, so that $\Sigma_3$
and $\Sigma_4$ join continuously at the maximum of the ring. Component~1 is an
inner disk of 0.3~$\mu$m grains. Component~2 is a disk of 300~$\mu$m grains.
Component~3 is a detached ring of 0.3~$\mu$m grains, and component~4 is its outer
edge, which consists of 0.5~$\mu$m grains (Section~\ref{sec:images}). The
components of submicron grains have $h_0=0.1$, whereas the large-grain disk is
geometrically thin with $h_0=0.01$, as expected for settled large grains.

\begin{figure*}
\largeplotthree{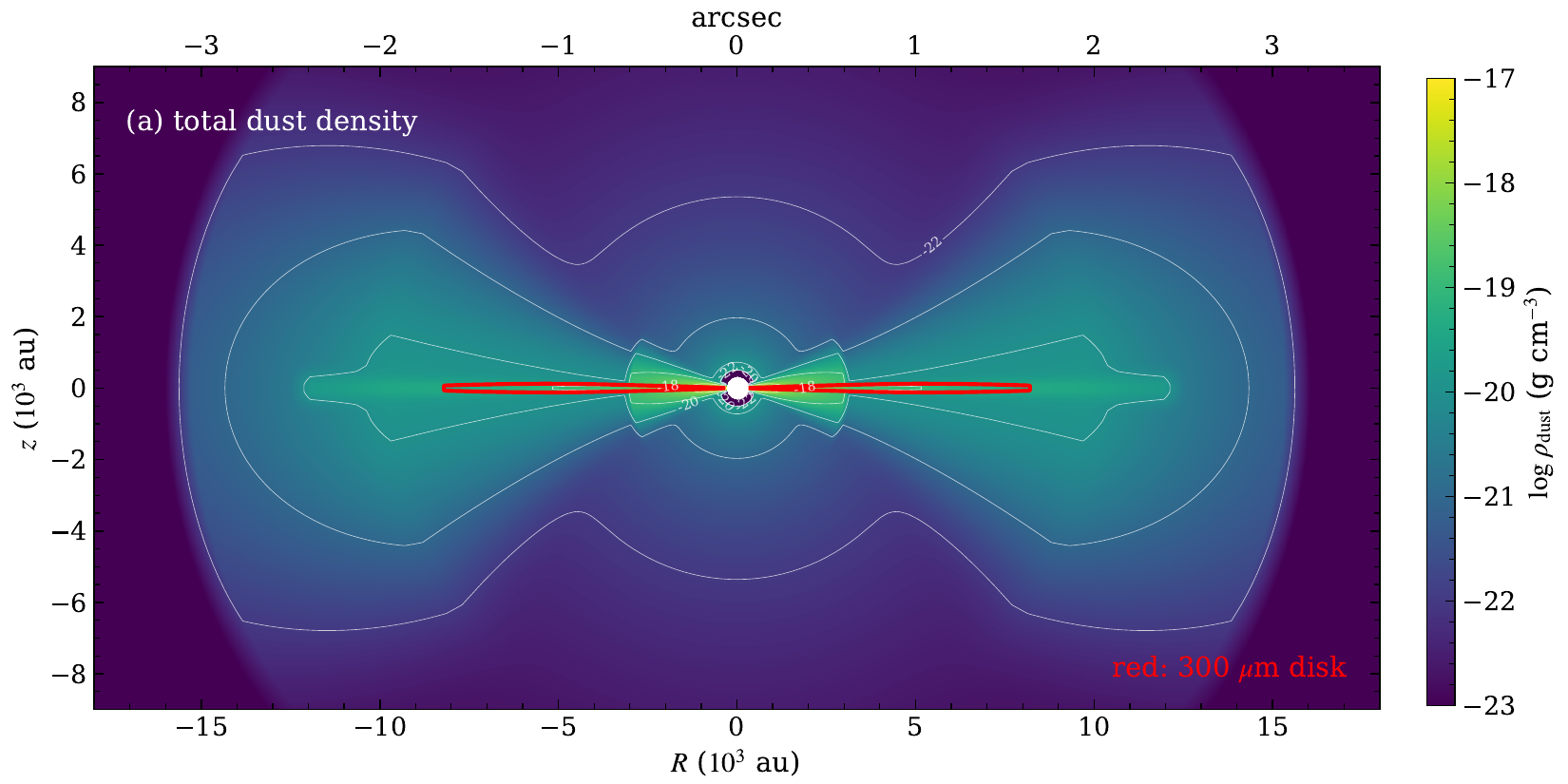}{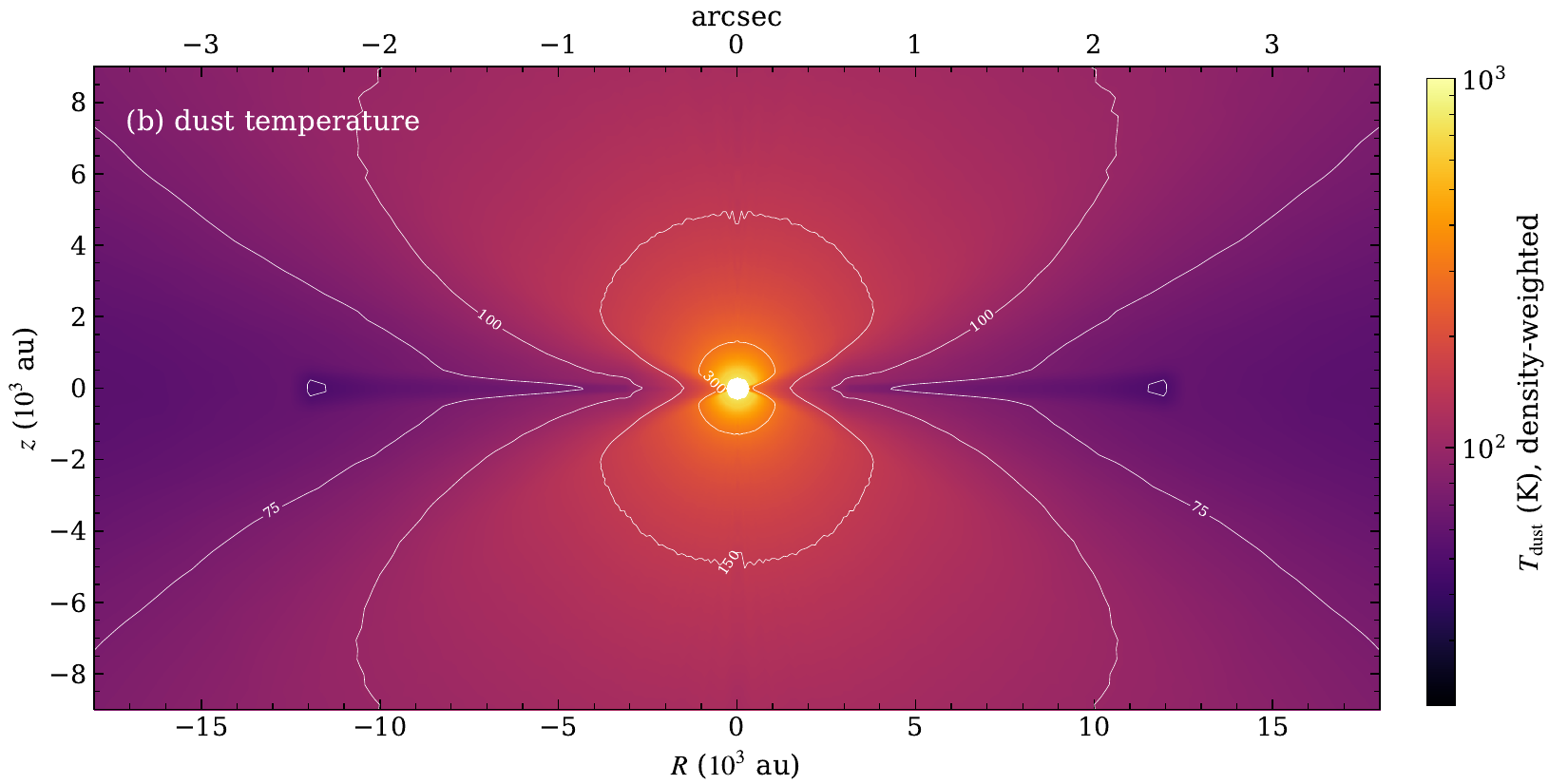}{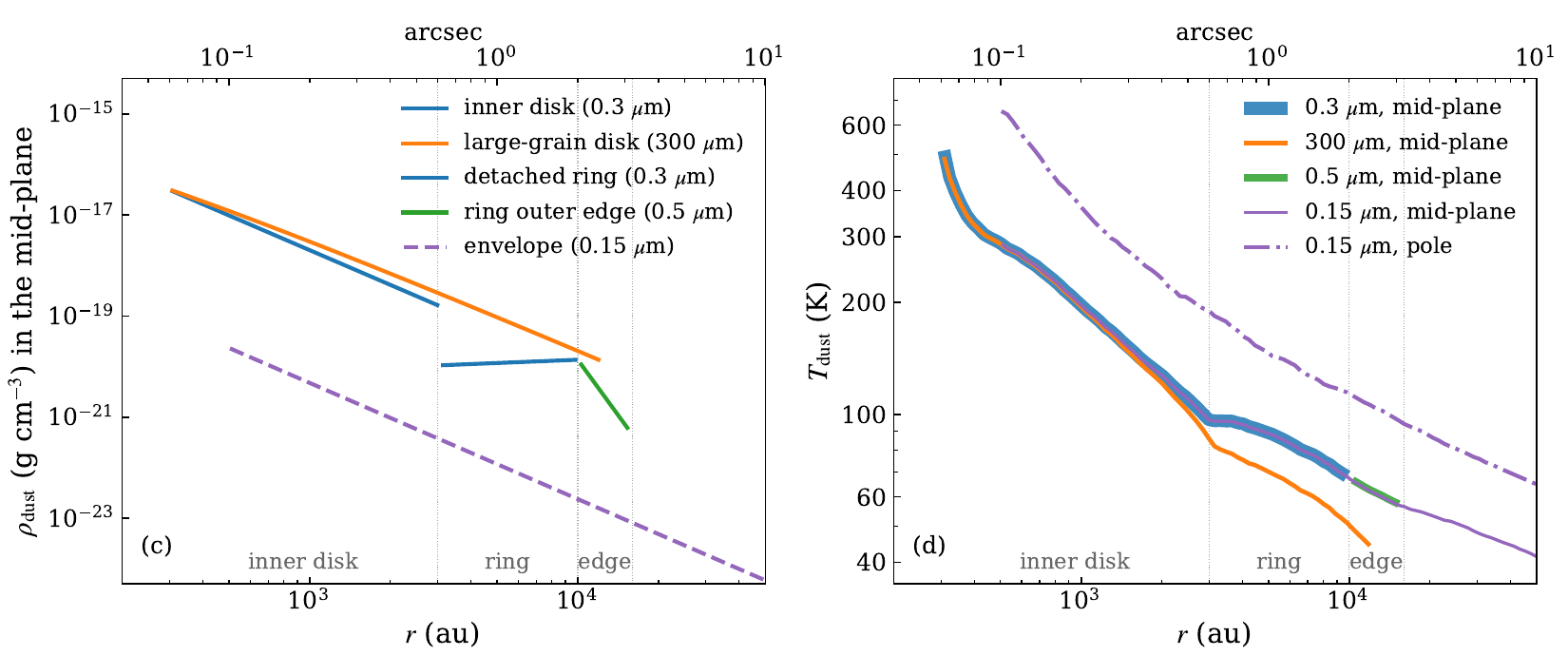}
\caption{Edge-on view of the adopted model: the dust distribution in the
meridional ($R$--$z$) plane, with the symmetry axis vertical. The system is
observed pole-on, i.e., along the $z$ axis from the top of panels (a) and (b).
(a) Dust density; white contours are spaced by 1~dex, and the red contour marks
the thin disk of 300~$\mu$m grains. (b) Dust temperature, density-weighted over
the grain populations, with isotherms. (c) Density of each component in the
equatorial plane. (d) Temperature of each grain population in the equatorial
plane (solid) and toward the pole (dashed-dot). Colors indicate the grain size and
are the same in panels (c) and (d): 0.3~$\mu$m (blue), 0.5~$\mu$m (green),
300~$\mu$m (orange), and 0.15~$\mu$m (purple). The mid-plane temperatures of
the 0.15, 0.3, and 0.5~$\mu$m grains are nearly identical; their curves are
drawn with different line widths. Dotted vertical lines in (c) and (d) mark the
boundaries between the inner disk, the ring, and its outer edge. The top axes
give the angular scale for $d=5$~kpc.
\label{fig:rz}}
\end{figure*}

The resulting surface density of the submicron grains decreases as $r^{-1}$ out
to 3000~au ($0\farcs6$ at 5~kpc). It then drops by a factor of $\approx16$ at the
inner edge of the ring, rises as $r^{1.5}$ to a maximum at 10\,000~au
($2\farcs0$), and falls steeply as $r^{-6}$ (Figure~\ref{fig:rz}). The positive
slope of the ring reproduces the nearly flat polarized-intensity profile observed
between $0\farcs7$ and $1\farcs7$ \citep{Shenoy2015}, and the steep outer edge of
weakly polarizing 0.5~$\mu$m grains reproduces its decline beyond $\approx2''$.

\paragraph{Envelope}
The disks are embedded in a spherical envelope of 0.15~$\mu$m grains,
\begin{equation}
  \rho_{\rm env}(r) = \rho_{{\rm env},0}
  \left(\frac{r}{R_{\rm env}}\right)^{-2.3},
  \label{eq:env}
\end{equation}
for $500\le r\le 5\times10^{5}$~au, where $R_{\rm env}=500$~au and $\rho_{{\rm
env},0}=2.34\times10^{-20}$~g~cm$^{-3}$. This corresponds to a radial optical
depth of $\tau_V=4.9$ ($\tau_{10\,\mu{\rm m}}=0.24$). A density exponent steeper
than that of a constant-velocity wind ($-2$) is required to suppress both the
far-infrared excess and the scattered light beyond $2\farcs5$. The adopted
parameters and the dust mass of each component are listed in
Table~\ref{tab:density}.

\textit{Interpretation of the multi-component structure}---The piecewise
description above should be regarded as an azimuthally averaged representation of
a structured equatorial outflow rather than as literally distinct disks. Clumpy,
arc-like, and spiral structures are commonly reported in the ejecta of yellow
hypergiants and other evolved massive stars
\citep[e.g.,][]{Humphreys2007,CastroCarrizo2007}, and when averaged over azimuth
they appear as radial zones with different density laws. The equatorial material
is not static. Its surface layers are shaped by the stellar wind into the flared,
hourglass-like structure of Figure~\ref{fig:rz}, and the condensations in the
inner nebula move nearly in the plane of the sky with individual velocities that
differ from the general expansion \citep{Tiffany2010}. Near the mid-plane,
however, the material appears to be close to settled. The submicron grains form a
flared layer ($h_0=0.1$), whereas the 300\,$\mu$m grains are confined to a
geometrically thin layer ($h_0=0.01$), as expected if the large grains have grown
in the dense mid-plane and settled toward it. Grain growth to sizes of
$\sim$100\,$\mu$m and vertical settling require high densities, low relative
velocities, and long residence times, which are not available in a uniformly
expanding shell. The presence of the large-grain disk therefore suggests that the
mid-plane of the equatorial structure is quasi-stationary. The submicron grains
in the disk components are also larger than those in the diffuse envelope, and
the largest of them are found at the outer edge of the ring.

Long-lived equatorial reservoirs containing large grains are well established
around silicate carbon stars, where oxygen-rich dust is stored in a disk separate
from the carbon-rich outflow \citep{Morris1987,LloydEvans1990,Yamamura2000}.
Grains as large as $\sim$300\,$\mu$m are required to explain the far-infrared
emission of \textit{IRAS}~09425$-$6040 \citep{Molster2001,Suh2016}, and similar
disks with large grains are found around post-AGB binaries
\citep[e.g.,][]{deRuyter2006,Bujarrabal2013}. Although these systems are of much
lower mass and luminosity than IRC\,+10420, they show that grain growth to
sub-millimeter sizes and settling can occur in equatorial structures around
evolved stars. In those systems the disks are thought to be stabilized by a
binary companion. No companion is known for IRC\,+10420, and the origin of its
quasi-stationary mid-plane remains open.

\paragraph{Structure of the adopted model}
Figure~\ref{fig:rz} shows the adopted model in a meridional plane, i.e., as it
would appear edge-on; we view the system from the top of panels (a) and (b). The
dust density (panel a) is concentrated toward the equatorial plane. The inner
disk fills the region within 3000~au, and the detached ring flares to heights of
$|z|\approx3000$~au at $R\approx10^{4}$~au. Beyond the ring the density is set by
the spherical envelope. The large-grain disk (red contour) is confined to a thin
layer with $|z|\lesssim200$~au.

Along the equatorial plane (panel c), the density of the inner disk decreases as
$r^{-2.3}$ (for $\Sigma\propto r^{-1}$ and $H\propto r^{9/7}$), from
$3.0\times10^{-17}$~g~cm$^{-3}$ at 300~au to $1.8\times10^{-19}$~g~cm$^{-3}$ at
2800~au. At the inner edge of the ring it drops by a factor of $\approx16$, to
$1.1\times10^{-20}$~g~cm$^{-3}$. Within the ring the density then increases
slowly ($\rho\propto r^{+0.2}$) to $1.4\times10^{-20}$~g~cm$^{-3}$ at 10\,000~au,
and in the outer edge it falls steeply ($\rho\propto r^{-7.3}$) to
$6\times10^{-22}$~g~cm$^{-3}$ at 15\,000~au. The 300~$\mu$m disk has a mid-plane
density comparable to that of the inner disk ($3.1\times10^{-17}$~g~cm$^{-3}$ at
300~au) because of its small scale height. The envelope contributes
$2.3\times10^{-20}$~g~cm$^{-3}$ at 500~au and less than 2\% of the mid-plane
density of the disk components at all radii inside 16\,000~au.

The dust temperature (panels b and d) is set by the absorption of stellar
radiation and its re-emission in the infrared. Along the pole, where the dust is
optically thin, the envelope grains are heated directly by the star. They reach
$\approx650$~K at 500~au, $\approx180$~K at 3000~au, and $\approx110$~K at
$10^{4}$~au.

The equatorial plane is shielded by the inner disk (the equatorial optical depth
at 10~$\mu$m is $\approx230$) and is heated mainly by the reprocessed infrared
radiation, so it is cooler. The mid-plane temperature decreases from 510~K at the
inner rim (300~au) to 200~K at 1000~au and 100~K at 2800~au. It remains nearly
constant (96--94~K) between 3100 and 4000~au, where the lower density at the
inner edge of the ring lets more radiation through, and then decreases to 68~K at
$10^{4}$~au and 41~K at $5\times10^{4}$~au.

In the optically thick mid-plane the 0.15, 0.3, and 0.5~$\mu$m grains have the
same temperature to within 1~K, although the temperature of each grain population
is computed separately. The radiation field there is dominated by the reprocessed
infrared emission and is close to a local blackbody, in which all grains reach
the same temperature regardless of their opacities. The remaining departures from
a blackbody affect the three submicron populations in the same way, because their
absorption efficiencies have the same wavelength dependence at infrared
wavelengths. The 300~$\mu$m grains absorb and emit like gray bodies. They have
the same temperature as the small grains within 1000~au, but are increasingly
cooler farther out, where the optical depth decreases: by 5\% at 2000~au, 15\% at
3100~au (82~K), and 29\% at 12\,000~au (44~K).

\subsection{Interstellar extinction} \label{sec:extinction}

For IRC\,+10420 we adopt an interstellar extinction of $A_V=5$~mag with
$R_V=3.1$, as adopted by \citet{Blocker1999} and \citet{DinhVTrung2009}, which is
also consistent with the DIB-based estimate of the interstellar color excess,
$E(B-V)_{\rm ISM}=1.4\pm0.5$ \citep{Oudmaijer1998}. We correct the observed
fluxes for it,
\begin{equation}
  F_\lambda^{\rm corr} = F_\lambda^{\rm obs}\,10^{0.4A_V k(\lambda)},
  \qquad k(\lambda)\equiv A_\lambda/A_V .
  \label{eq:dered}
\end{equation}
For $\lambda\le2.2~\mu$m we use the extinction law of \citet{Cardelli1989}.
Between 2.2 and 8~$\mu$m we use the near-infrared law of \citet{Indebetouw2005},
\begin{equation}
  k(\lambda)=k_K\left[0.43+0.57\left(\frac{\lambda}{2.2~\mu{\rm m}}\right)^{-3}\right],
  \label{eq:nir}
\end{equation}
where $k_K$ is the \citet{Cardelli1989} value at 2.2~$\mu$m. Beyond 8~$\mu$m the
extinction is the sum of a continuum, $0.43\,k_K(\lambda/8~\mu{\rm m})^{-2}$, and
the interstellar silicate features at 9.7 and 18~$\mu$m, represented by
peak-normalized Drude profiles with widths of 2.5 and 6~$\mu$m, a relative
strength of 0.4 for the 18~$\mu$m feature, and $\tau_{9.7}/A_V=1/18.5$. The model
fluxes are intrinsic, i.e., they include the circumstellar extinction computed by
the radiative transfer but no interstellar extinction. The uncertainty of $A_V$
($\pm0.5$~mag) is propagated into the flux errors of the dereddened photometry.

The circumstellar envelope contributes $A_V^{\rm cs}=1.85$\,mag and $E(B-V)^{\rm
cs}=0.46$ to the line of sight. These values are measured by comparing the model
SED, which includes the light scattered back into the beam, with the bare stellar
spectrum. The total color excess of the model is therefore
$E(B-V)=A_V/3.1+E(B-V)^{\rm cs}=2.08$. Of this, 1.61 is interstellar, in
agreement with the DIB estimate. For comparison, \citet{Oudmaijer1996} derived a
total color excess of $E(B-V)=2.08$ by fitting the observed SED with a 6000\,K
photospheric model. This value includes both the interstellar and the
circumstellar reddening. Because a 7000\,K photosphere is intrinsically bluer by
0.2--0.3\,mag in $B-V$, the same fit would give $E(B-V)\approx2.3$ for the
stellar temperature adopted here. The remaining difference of $\lesssim0.25$\,mag
is comparable to the variability of the star in the $B$ band. The total
extinction toward the star is therefore $A_V\approx6.85$\,mag, of which 5\,mag is
interstellar and $\approx1.85$\,mag is circumstellar.

For $T_\star=7000$ K, the color excess constrains the size of the envelope
grains. We replaced the 0.15~$\mu$m envelope grains by grains with $a_0=0.1$ or
0.2~$\mu$m (same $\sigma_a$), keeping the disks unchanged and varying the
envelope density by factors of 0.7--1.3. With 0.2\,$\mu$m grains, whose
extinction is nearly gray ($\kappa^{\rm ext}_B/\kappa^{\rm ext}_V=1.10$), the
best SED fit ($\sigma_{\log}=0.071$\,dex) gives a total color excess of only
1.82, below the SED-based estimate. With 0.1\,$\mu$m grains ($\kappa^{\rm
ext}_B/\kappa^{\rm ext}_V=1.62$), the SED requires a denser envelope, and even
then the fit is worse ($\sigma_{\log}=0.075$\,dex) and the color excess too large
(2.40). The 0.15~$\mu$m grains, whose extinction curve resembles that of
interstellar dust, give both the best SED fit and a total color excess consistent
with the SED-based estimate, while keeping the interstellar part consistent with
the DIB measurement. For hotter stars, smaller grains are favored
(Section~\ref{sec:sed}).

\begin{figure*}
\plotone{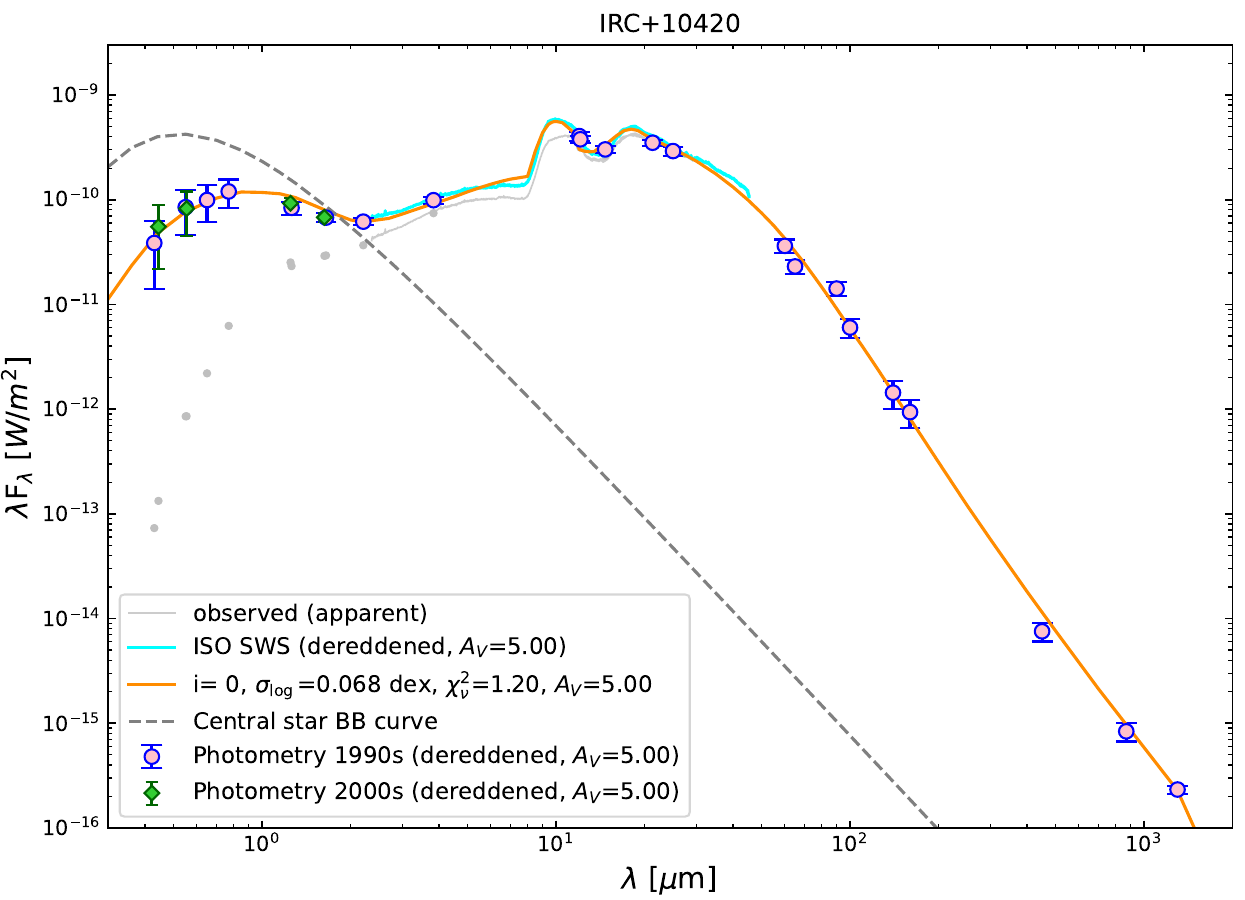}
\caption{Spectral energy distribution of IRC\,+10420 and the adopted model.
Gray symbols and the gray line show the observed (apparent) photometry and
ISO/SWS spectrum. Open circles and the cyan line show the 1990s photometry and
the ISO/SWS spectrum (1996--97) corrected for an interstellar extinction of
$A_V=5$~mag with $R_V=3.1$ (Section~\ref{sec:extinction}). Filled diamonds show the
unsaturated photometry of the 2000s (APASS $B$ and $V$, 2MASS $J$ and $H$),
dereddened in the same way. Error bars include the uncertainty of the
interstellar extinction ($\pm0.5$~mag in $A_V$). All data are used in the fit.
The orange line is the adopted model seen pole-on ($i=0^\circ$), which includes
the circumstellar extinction and the scattered light but no interstellar
extinction ($\sigma_{\log}=0.068$~dex, $\chi^2_\nu=1.20$;
Equation~\ref{eq:siglog}). The dashed line is the unattenuated central star, a
7000~K blackbody with $L_\star=4.5\times10^{5}\,L_\odot$ at $d=5$~kpc.
\label{fig:sed}}
\end{figure*}

\section{Comparison with the observations} \label{sec:results}

\subsection{The SED} \label{sec:sed}

Figure~\ref{fig:sed} compares the observed SED with the model SED. We measure the
agreement between the model and the dereddened SED with the weighted
root-mean-square logarithmic residual,
\begin{equation}
  \sigma_{\log} = \left[\frac{\sum_i w_i\,
  (\log F_i^{\rm obs}-\log F_i^{\rm mod})^2}{\sum_i w_i}\right]^{1/2},
  \label{eq:siglog}
\end{equation}
where the weights $w_i$ give equal total weight to the photometry and to the
ISO/SWS spectrum. The photometry comprises the 1990s data set, which is
contemporaneous with the ISO/SWS spectrum, and unsaturated photometry from the
2000s ($B$ and $V$ from APASS and $J$ and $H$ from 2MASS); saturated measurements
(2MASS $K_s$, WISE) and the very broad Gaia bands are not used. We also compute
the reduced $\chi^2_\nu$ with an error floor of 10\%, and the scale factor $s$
that minimizes $\sigma_{\log}$ when the model fluxes are multiplied by a
constant. A value of $s\neq1$ indicates that the luminosity should be changed by
the same factor (at a fixed distance).

The model SED, computed at 91 wavelengths with $10^6$ scattering photon packages
at each wavelength, is interpolated to the observed wavelengths with a monotonic
cubic (PCHIP) interpolation in $\log\lambda$--$\log F_\lambda$. The reduced
$\chi^2_\nu$ uses the same weights as $\sigma_{\log}$, with the effective number
of data points $N_{\rm eff}=(\sum w_i)^2/\sum w_i^2$.

The adopted model (Table~\ref{tab:density}) reproduces the SED with
$\sigma_{\log}=0.068$~dex, $\chi^2_\nu=1.20$, and $s=1.01$
(Figure~\ref{fig:sed}); the 2000s photometry agrees with the 1990s photometry
within 0.11~dex. The largest residuals, up to 0.19~dex, occur in the far-infrared
and submillimeter, where the data come from different instruments and epochs;
part of the excess at 450\,$\mu$m (0.04\,dex) is due to emission outside the
SCUBA beam (Section~\ref{sec:obs_sed}).

\begin{deluxetable}{cccccccc}
\tablecaption{Dependence on the Stellar Temperature\label{tab:tstar}} \tablehead{
\colhead{$T_\star$} & \colhead{$a_0^{\rm env}$} & \colhead{$f_{\rm env}$} &
\colhead{$L_\star$} & \colhead{$A_V^{\rm cs}$} & \colhead{$E(B-V)^{\rm cs}$} &
\colhead{$\sigma_{\log}$} & \colhead{$\chi^2_\nu$} \\
\colhead{(K)} & \colhead{($\mu$m)} & \colhead{} & \colhead{($10^5\,L_\odot$)} &
\colhead{(mag)} & \colhead{(mag)} & \colhead{(dex)} & \colhead{}} \startdata
7000\tablenotemark{a} & 0.15 & 1.00 & 4.56 & 1.85 & 0.46 & 0.068 & 1.20 \\
8000 & 0.10 & 1.15 & 4.56 & 1.60 & 0.69 & 0.067 & 1.24 \\
8000 & 0.15 & 1.00 & 4.62 & 1.85 & 0.46 & 0.068 & 1.19 \\
8000 & 0.20 & 1.00 & 4.69 & 1.92 & 0.21 & 0.085 & 1.38 \\
8500 & 0.10 & 1.15 & 4.59 & 1.60 & 0.69 & 0.069 & 1.34 \\
8500 & 0.15 & 1.00 & 4.65 & 1.85 & 0.47 & 0.076 & 1.38 \\
8500 & 0.20 & 0.85 & 4.69 & 1.58 & 0.17 & 0.096 & 1.56 \\
\enddata
\tablecomments{Best-fitting envelope density factor $f_{\rm env}$ (relative to
the adopted model) for each $T_\star$ and envelope grain size. All models have
$A_V^{\rm ISM}=5$\,mag, $R_V=3.1$, $d=5$\,kpc, $i=0^\circ$, and the same disk
components. $L_\star=s\times4.5\times10^5\,L_\odot$.} \tablenotetext{a}{Adopted
model.}
\end{deluxetable}

\textit{Dependence on the stellar temperature}---\label{sec:tstar}
Table~\ref{tab:tstar} compares the adopted model with models of $T_\star=8000$
and 8500\,K. For each temperature, we varied the envelope grains ($a_0=0.10$,
0.15, and 0.20\,$\mu$m) and the envelope density (by factors of 0.7--1.3). We
refitted $L_\star$ through the scale factor $s$ and kept the disk components
unchanged. The comparison gives three results.
\begin{enumerate}[leftmargin=1em]
\setlength{\itemsep}{1pt}
\item The SED does not distinguish between these temperatures. The best
    models reach $\sigma_{\log}=0.067$--0.069\,dex at all three temperatures.
    The spectroscopic temperatures of 7900--8500\,K
    \citep{Klochkova1997,Nieuwenhuijzen2000} are therefore fully compatible
    with the data.
\item The circumstellar color excess is set by the envelope grains and is
    independent of $T_\star$. It is $E(B-V)^{\rm cs}=0.46$ for 0.15\,$\mu$m
    grains and 0.69 for 0.10\,$\mu$m grains. A hotter, intrinsically bluer star
    therefore requires smaller grains. At 8000\,K, 0.10 and 0.15\,$\mu$m grains
    fit equally well. At 8500\,K, 0.10\,$\mu$m grains are preferred
    ($\sigma_{\log}=0.069$ vs.\ 0.076\,dex). Grains of 0.20\,$\mu$m give poorer
    fits at all temperatures ($\sigma_{\log}\geq0.085$\,dex). The stellar
    temperature and the envelope grain size are thus degenerate, and the SED
    constrains the envelope grains to $a_0\approx0.10$--0.15\,$\mu$m. The total
    color excess of the models with 0.10\,$\mu$m grains, $E(B-V)=2.31$, agrees
    with the SED-based estimate corrected for the intrinsic color of a hotter
    star (Section~\ref{sec:extinction}).
\item The luminosity is insensitive to the stellar temperature. The best models
    have $L_\star=4.6\times10^5\,L_\odot$, and the circumstellar extinction is
    $A_V^{\rm cs}=1.6$--1.85\,mag.
\end{enumerate}
We retain $T_\star=7000$\,K for the adopted model. None of our conclusions on the
dust distribution depend on this choice.

\paragraph{Dependence on the extinction assumption}
If instead the entire selective extinction is attributed to the interstellar
medium ($A_V=6.5$~mag, which corresponds to the total color excess of
\citet{Oudmaijer1996} for $R_V=3.1$) and the envelope consists of 0.3~$\mu$m
grains that produce an almost gray circumstellar extinction, the SED is fitted
about equally well ($\sigma_{\log}=0.077$~dex) with
$L_\star=6.0\times10^{5}\,L_\odot$. The luminosity is higher in this case because
starlight absorbed by interstellar dust is lost from the SED, whereas starlight
absorbed by circumstellar dust is re-emitted in the infrared. The derived
luminosity therefore depends on how the total extinction of $\approx7$~mag is
divided between the interstellar medium and the circumstellar envelope, and the
range $L_\star=(4.5$--$6.0)\times10^{5}\,L_\odot$ reflects this systematic
uncertainty. We prefer the adopted model because its interstellar extinction
agrees with the independent DIB estimate \citep{Oudmaijer1998}, reproduces the
observed color excess with a single envelope grain population, and reproduces the
polarized light within $1\farcs7$ better, since the 0.15~$\mu$m envelope grains
scatter little at $K'$.

\textit{Luminosity and viewing angle}---The model is axisymmetric, so the
bolometric flux received by an observer depends on the viewing angle.
Table~\ref{tab:fbol} lists the apparent luminosity, $L_{\rm app}(i)\equiv4\pi
d^2F_{\rm bol}(i)$, of the adopted model. Seen pole-on, $L_{\rm
app}=1.55\,L_\star$. The equatorial components absorb the starlight emitted
toward low latitudes and re-emit and scatter much of it toward the poles. A
pole-on observer also sees the star through the tenuous polar envelope ($A_V^{\rm
cs}\approx1.85$\,mag). Near the equatorial plane, $L_{\rm app}$ falls to
$\approx0.2\,L_\star$. The average of $L_{\rm app}$ over $\cos i$ equals the
total model luminosity to within 4\%, which confirms energy conservation in the
calculation. For $i\lesssim50^\circ$, a luminosity derived by integrating the SED
under the assumption of isotropic emission overestimates $L_\star$ by a factor of
1.3--1.6. The same effect was found for the dusty torus of WOH G64
\citep{Ohnaka2008}.

\begin{figure}
\smallplot{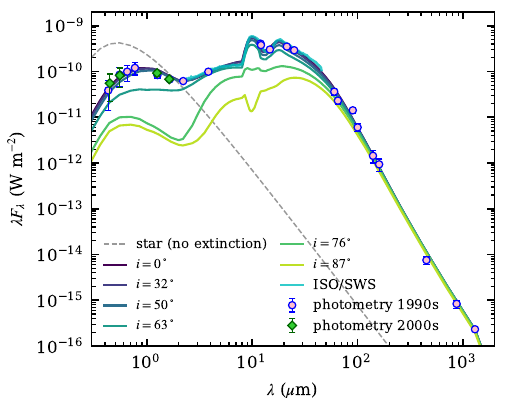}
\caption{SEDs of the adopted model seen at the inclinations of
Table~\ref{tab:fbol} ($i=0^\circ$, $32^\circ$, $50^\circ$, $63^\circ$,
$76^\circ$, and $87^\circ$), compared with the dereddened photometry of the 1990s
(open circles) and the 2000s (filled diamonds) and the ISO/SWS spectrum (as in
Figure~\ref{fig:sed}). The dashed gray line is the unattenuated central star. For
$i\lesssim50^\circ$ the SED changes little. At larger inclinations the star is
obscured by the equatorial components: the optical and near-infrared flux drops
by more than an order of magnitude at $i=87^\circ$, and the $9.7\,\mu$m silicate
feature turns from emission into absorption. The fluxes beyond $\approx60\,\mu$m
are nearly independent of inclination.}
\label{fig:sedincl}
\end{figure}

Figure~\ref{fig:sedincl} shows the corresponding SEDs. Seen close to edge-on, the
same object would appear as a much redder source with the silicate feature in
absorption, while its far-infrared and submillimeter fluxes would be nearly
unchanged. The fit to the observed SED deteriorates with inclination:
$\sigma_{\log}=0.068$, 0.075, 0.105, and 0.22\,dex for $i=0^\circ$, $32^\circ$,
$50^\circ$, and $63^\circ$, and the scale factor needed to match the observed
flux level increases from $s=1.01$ to 1.05, 1.15, and 1.50. The SED itself
therefore favors $i\lesssim30^\circ$, consistent with the pole-on view inferred
from the kinematics.

\begin{deluxetable*}{lccccccccccc}
\tabletypesize{\small} \tablecaption{Apparent Luminosity of the Adopted
Model\label{tab:fbol}} \tablewidth{0pt} \tablehead{\colhead{$i$ (deg)} &
\colhead{0} & \colhead{18.2} & \colhead{31.8} & \colhead{41.4} & \colhead{49.5} &
\colhead{56.6} & \colhead{63.3} & \colhead{69.5} & \colhead{75.5} &
\colhead{81.4} & \colhead{87.1}} \startdata
$L_{\rm app}/L_\star$ & 1.55 & 1.54 & 1.48 & 1.41 & 1.33 & 1.20 & 0.99 & 0.74 & 0.49 & 0.29 & 0.22 \\
\enddata
\tablecomments{$L_{\rm app}=4\pi d^2F_{\rm bol}(i)$ for the adopted model
($L_\star=4.5\times10^5\,L_\odot$). The inclinations other than $i=0^\circ$ are
uniformly spaced in $\cos i$; their mean is $0.97\,L_\star$.}
\end{deluxetable*}

\textit{Comparison with previous luminosity estimates}---Integrated in this way,
the dereddened SED gives $L_{\rm app}\approx7\times10^5\,L_\odot$ at 5\,kpc.
\citet{Oudmaijer2013} obtained a similar value, $L=25\,462\,(d/{\rm
kpc})^2\,L_\odot$, i.e., $6.4\times10^5\,L_\odot$ at 5\,kpc and
$4.6\times10^5\,L_\odot$ at the \textit{Gaia} distance of 4.26\,kpc
\citep{Koumpia2022}. The difference from our value,
$L_\star=4.5\times10^5\,L_\odot$ at 5\,kpc, therefore arises mainly from the
anisotropy of the emission, not from the fit to the SED. A second, smaller
systematic effect comes from the division of the extinction. Attributing all the
selective extinction to the interstellar medium raises $L_\star$ to
$6.0\times10^5\,L_\odot$ (see above). At the \textit{Gaia} distance, the adopted
model corresponds to $L_\star\approx3.3\times10^5\,L_\odot$.

\subsection{$K'$ polarimetric images} \label{sec:images}

To compare with the MMT-Pol $K'$ ($2.2\,\mu$m) observations \citep{Shenoy2015},
we compute Stokes $I$, $Q$, and $U$ images at 2.2~$\mu$m with a pixel size of
$0\farcs01$. We convolve them separately with an effective point spread function
(PSF), $P_{\rm eff}=P_{\rm M}\ast G$, where $G$ is a Gaussian with a full width
at half maximum (FWHM) of $0\farcs2$, which represents the smoothing applied in
the data reduction, and $P_{\rm M}$ is a Moffat profile
\citep{Moffat1969,Trujillo2001},
\begin{equation}
  P_{\rm M}(r)=\frac{\beta-1}{\pi r_c^{2}}
  \left[1+\left(\frac{r}{r_c}\right)^{2}\right]^{-\beta},
  \label{eq:moffat}
\end{equation}
with FWHM $=2r_c(2^{1/\beta}-1)^{1/2}=0\farcs2$ and $\beta=1.6$
($r_c=0\farcs136$), which reproduces the observed stellar profile. The effective
PSF has a FWHM of $0\farcs33$, and 20\% of its light lies beyond $0\farcs5$. The
convolved images are resampled to the detector pixel size ($0\farcs043$). The
polarized intensity $I_P=(Q^2+U^2)^{1/2}$ and the fractional polarization $p^{\rm
raw}=I_P/I^{\rm raw}$ are computed from the convolved Stokes parameters. The
profiles are averaged in $0\farcs2$ annuli within the east and west sectors, as
in the observations.

The stellar contribution is separated within the same Monte Carlo calculation.
The unresolved star is the central excess ($r<0\farcs05$, inside the inner radius
of the dust) of the image computed with the star over the image computed without
it. This excess includes the circumstellar extinction of the direct stellar beam.
Its convolution with $P_{\rm eff}$ gives the stellar profile $I^\star$, and the
nebular intensity is $I^{\rm neb}=I^{\rm raw}-I^\star$. The Monte Carlo
uncertainties of the annulus means, estimated from two independent calculations
and from the difference between the east and west profiles, are below 1\%.
Changing the FWHM of $G$ between 0 and $0\farcs2$ changes the comparison metrics
below by less than 0.004~dex.

We compare the models with the observed profiles between $0\farcs7$ and
$2\farcs9$. The absolute calibration of the observed intensities is not used: all
intensities are normalized by $I^{\rm raw}$ at $0\farcs7$. As metrics we use the
rms logarithmic residuals of the shape of $I^{\rm raw}$, the shape of $I_P$, the
ratio $I_P/I^\star$ (with $I^\star$ defined as in \citealt{Shenoy2015}, i.e., the
PSF scaled to $I^{\rm raw}$ at $0\farcs5$), and $p^{\rm raw}$.

\begin{figure}
\smallplot{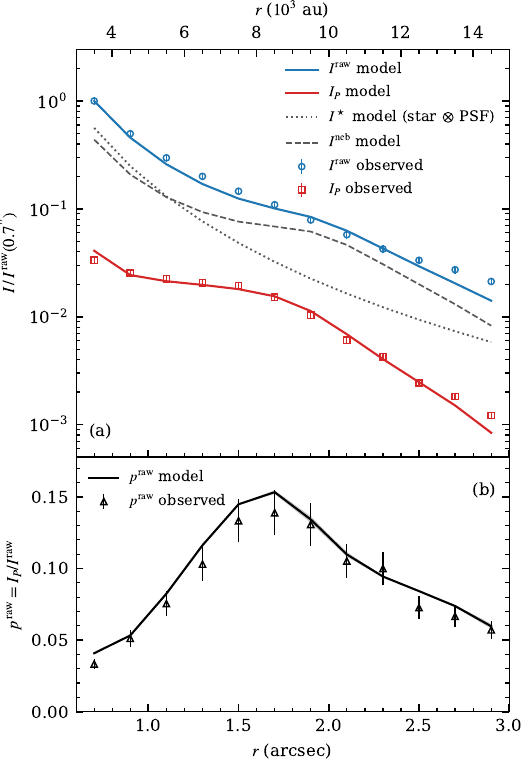}
\caption{Radial profiles of the $K'$ ($2.2\,\mu$m) emission of IRC\,+10420 compared with
the adopted model. (a) Total intensity $I^{\rm raw}$ (blue) and polarized
intensity $I_P$ (red), each normalized by $I^{\rm raw}$ at $0\farcs7$.
Symbols are the observations \citep{Shenoy2015}; lines are the model
convolved with the effective PSF (Moffat, FWHM $0\farcs2$, $\beta=1.6$,
convolved with a $0\farcs2$ Gaussian). The dotted and dashed gray lines are
the stellar ($I^\star$) and nebular ($I^{\rm neb}$) contributions of the
model. (b) Fractional polarization $p^{\rm raw}=I_P/I^{\rm raw}$. Shaded
bands show the Monte Carlo uncertainties of the model (mostly smaller than
the line width). The top axis gives the projected distance for $d=5$~kpc.
\label{fig:sbprofile}}
\end{figure}

Figure~\ref{fig:sbprofile} compares the model with the $K'$ profiles. The rms
residuals are 0.076\,dex for the shape of $I^{\rm raw}$, 0.061\,dex for the shape
of $I_P$, 0.056\,dex for $I_P/I^\star$, and 0.044\,dex for $p^{\rm raw}$.
Scattered light contributes 44\%--74\% of $I^{\rm raw}$ between $0\farcs5$ and
$2\farcs3$, so the extended profile, which is shallower than the PSF wings, is
produced by the nebula. The flat polarized-intensity profile between $0\farcs7$
and $1\farcs7$ arises from the rising surface density of the ring
(Equation~\ref{eq:s3}), and its decline beyond $2''$ from the outer edge of
$0.5\,\mu$m grains. The fractional polarization peaks at 0.15 at $1\farcs7$
(observed: 0.14). The largest deviations occur beyond $2\farcs5$, where the model
$I^{\rm raw}$ is 12\%--33\% low.

\begin{figure*}
\largeplotone{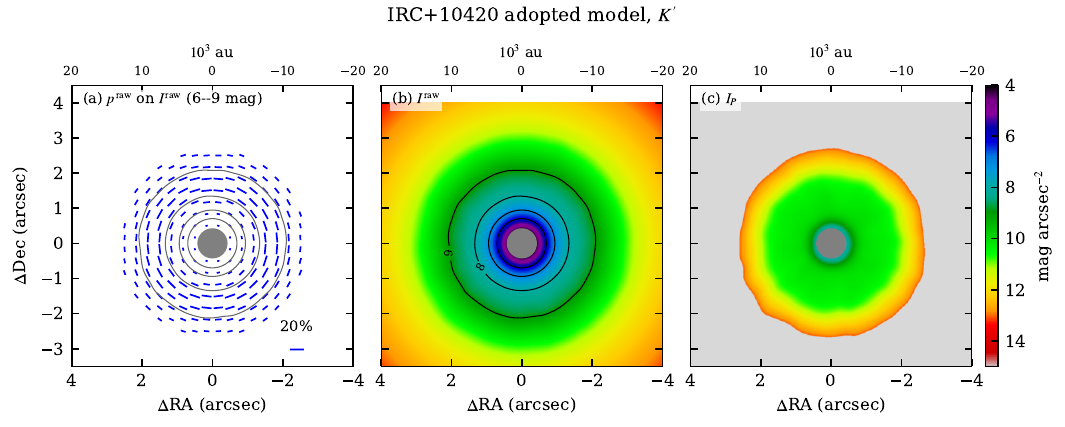}
\caption{Model $K'$ ($2.2\,\mu$m) images of IRC\,+10420 displayed as in Figure~1 of
\citet{Shenoy2015}, convolved with the effective PSF (Moffat, FWHM
$0\farcs2$, $\beta=1.6$, convolved with a $0\farcs2$ Gaussian).
(a) Polarization vectors ($p^{\rm raw}$, binned to $0\farcs3$) on contours
of $I^{\rm raw}$ at 6, 7, 8, and 9~mag~arcsec$^{-2}$; the bar indicates
20\% polarization. (b) Total intensity $I^{\rm raw}$ and (c) polarized
intensity $I_P$ in mag~arcsec$^{-2}$. The central $0\farcs4$, which is
saturated in the observations, is masked, and regions with $I_P$ fainter
than 13~mag~arcsec$^{-2}$, where the Monte Carlo noise is large, are not
shown. The top axes give the projected distance for $d=5$~kpc.
\label{fig:fig1style}}
\end{figure*}

Figure~\ref{fig:fig1style} shows the model images in the format of Figure~1 of
\citet{Shenoy2015}. The isophotes of $I^{\rm raw}$ are nearly circular, as
expected for a pole-on system, and the polarized intensity forms a broad plateau
out to $\approx2''$, bounded by the steep outer edge of the ring. The
polarization vectors are centrosymmetric throughout: their orientations deviate
from the tangential direction by less than $10^\circ$ (median $1^\circ$),
including the outer edge of 0.5~$\mu$m grains, as in the observed vector map. In
$0\farcs3$ bins, the fractional polarization rises from 2--5\% at
$0\farcs4$--$1''$ to a maximum of 16\% at $1\farcs7$ and decreases to 4--9\% at
$2\farcs5$--$3''$. The bin maximum is slightly higher than the peak of the
azimuthally averaged profile (0.15), and it depends on the binning by about one
percentage point.

\paragraph{Grain size in the outer ring}
Table~\ref{tab:ringtest} compares models with different grains in the ring and
its outer edge. All fit the SED equally well ($\sigma_{\log}=0.067$--0.071\,dex),
so the grain size is constrained by the polarimetry alone. With $0.3\,\mu$m
grains also in the outer edge, the polarized intensity beyond $1\farcs9$ is
1.2--2.1 times too high and $p^{\rm raw}$ peaks at 0.19--0.21 at
$1\farcs9$--$2\farcs1$; with $0.5\,\mu$m grains throughout the ring,
$I_P/I^\star$ is $\approx30\%$ too low within $1\farcs7$ and $p^{\rm raw}$ only
$\approx0.09$. Moving the transition to 9000\,au makes the polarized intensity
beyond $1\farcs7$ 20\%--50\% too low, and moving it to 11\,000\,au makes it
1.5--2.0 times too high, so the transition lies close to 10\,000\,au ($2''$).
Grains with $a_0=0.7$ or $1\,\mu$m polarize light scattered through
$\approx90^\circ$ negatively or hardly at all ($-Z_{12}/Z_{11}$ between $-0.37$
and 0; Figure~\ref{fig:polsize}), and would produce radial vectors or negligible
polarization instead of the observed centrosymmetric pattern.

\begin{deluxetable*}{lcccccc}
\tablecaption{Tests of the Grain Size in the Detached Ring\label{tab:ringtest}}
\tablehead{ \colhead{Ring / edge grains} & \colhead{$R_b$} & \colhead{SED} &
\colhead{$I_P/I^\star$} & \colhead{$I^{\rm raw}$} & \colhead{$p^{\rm raw}$} &
\colhead{$p^{\rm raw}_{\rm max}$} \\
\colhead{($\mu$m)} & \colhead{(au)} & \colhead{(dex)} & \colhead{(dex)} &
\colhead{(dex)} & \colhead{(dex)} & \colhead{}} \startdata
0.3 / 0.5 (adopted) & 10\,000 & 0.068 & 0.056 & 0.076 & 0.044 & 0.15 ($1\farcs7$) \\
0.3 / 0.5           &  9000   & 0.071 & 0.144 & 0.110 & 0.077 & 0.13 ($1\farcs5$) \\
0.3 / 0.5           & 11\,000 & 0.069 & 0.154 & 0.071 & 0.102 & 0.18 ($1\farcs7$) \\
0.3 / 0.3           & 10\,000 & 0.068 & 0.084 & 0.139 & 0.169 & 0.19 ($1\farcs9$) \\
0.3 / 0.3           & 11\,000 & 0.069 & 0.178 & 0.103 & 0.230 & 0.21 ($2\farcs1$) \\
0.5 / 0.5           & 10\,000 & 0.067 & 0.131 & 0.081 & 0.144 & 0.09 ($1\farcs7$) \\
\enddata
\tablecomments{$R_b$ is the radius of the transition between the ring and its
outer edge; the surface density of the ring is unchanged, and the outer edge
joins it continuously at $R_b$. Columns 3--6 give the rms logarithmic residuals
of the SED, of $I_P/I^\star$, of the shape of $I^{\rm raw}$, and of $p^{\rm raw}$
(Section~\ref{sec:images}); the last column gives the peak of the azimuthally
averaged $p^{\rm raw}$ (observed: 0.14 at $1\farcs7$).}
\end{deluxetable*}

We also note that subtracting the stellar PSF scaled at $0\farcs5$, as in the
observational analysis, removes most of the nebular light close to the star.
Applied to the model, this procedure gives $I^{\rm neb}/I^{\rm raw}=0.01$ at
$0\farcs7$, whereas the true fraction is 0.44. The nebular intensities and
polarizations derived in this way within $\approx1''$ should therefore be
interpreted with caution.

\section{Discussion} \label{sec:discussion}

\subsection{Mass loss} \label{sec:massloss}

\textit{The envelope}---We assume that the envelope dust moves radially at a
constant expansion velocity, $v_{\rm exp}=35$\,km\,s$^{-1}$
\citep{Oudmaijer1996,CastroCarrizo2007}. The dust mass-loss rate is then $\dot
M_{\rm d}=4\pi r^2\rho_{\rm env}v_{\rm exp}\approx9\times10^{-7}\, (r/500\,{\rm
au})^{-0.3}\,M_\odot\,{\rm yr^{-1}}$. For a gas-to-dust mass ratio $\psi$, the
total rate is $\approx9\times10^{-5}\,(\psi/100)\,M_\odot$\, yr$^{-1}$ at 500\,au
and $\approx1\times10^{-5}\,(\psi/100)\,M_\odot$\, yr$^{-1}$ at
$5\times10^5$\,au, corresponding to flow times of $\approx70$ to
$\approx7\times10^4$\,yr. For red supergiants, $\psi$ is usually taken to be
100--200; $\psi=200$ corresponds to complete condensation of the refractory
elements at solar abundances \citep{Mauron2011}, and would double these rates.
The rates scale as $(\psi/100)(v_{\rm exp}/35\,{\rm km\,s^{-1}}) (d/5\,{\rm
kpc})$. Taken at face value, the density profile $\rho\propto r^{-2.3}$ implies
that the rate has increased toward the present. The scattered light, however,
constrains the envelope only out to $\approx2\farcs5$ ($1.2\times10^4$\,au);
farther out, the envelope is constrained only by the far-infrared and
submillimeter fluxes, which are insensitive to the density slope. This trend
should therefore be regarded as tentative.

\textit{The equatorial components}---The disk-like components contain
$6.3\times10^{-2}\,M_\odot$ of dust, i.e., $\approx6\,(\psi/100)\,M_\odot$ of gas
and dust, about six times the dust mass of the envelope. We do not convert their
density into a mass-loss history. The condensations in the equatorial plane move
with individual velocities that differ from the general expansion
\citep{Tiffany2010}, so neither a single expansion velocity nor a one-to-one
relation between radius and ejection time can be assumed. The radial structure of
the disk components---the dense inner disk, the density contrast at 3000\,au, and
the ring---may reflect variations in the mass-loss rate, but also deceleration,
accumulation, or interaction within the outflow. Its total mass nevertheless
shows that the equatorial outflow carries most of the circumstellar dust.

\textit{The ring and the wind}---The expansion velocity of the envelope, however,
provides a natural upper limit to the outward velocity of the equatorial
material, which is more likely to be decelerated than accelerated relative to the
wind. With $v\leq35$\,km\,s$^{-1}$, the outer edge of the ring (16\,000\,au), its
maximum ($10^4$\,au), and its inner edge (3000\,au) were ejected at least
$\approx2200$, 1350, and 400\,yr ago, respectively (for $d=5$\,kpc).
Correspondingly, the mass of the ring and its outer edge
($3.4\times10^{-2}\,M_\odot$ of dust) implies an equatorial mass-loss rate of at
most $\approx2\times10^{-3}\,(\psi/100)\,M_\odot$\,yr$^{-1}$ during its
formation. These limits are consistent with the high mass-loss phase that ended
$\sim2000$\,yr ago according to mid- and far-infrared imaging \citep{Shenoy2016}.
We therefore suggest that the ring consists of the equatorial material of this
phase, which has since been decelerated, while the faster, spherical wind
continued to expand. The flared surface of the ring and the low-density polar
regions (Figure~\ref{fig:rz}) are consistent with a fast wind sweeping the
surface layers of the slower equatorial material toward the poles, while the
dense mid-plane, where the large grains have grown and settled, remains close to
the equatorial plane.

\subsection{Comparison with previous studies} \label{sec:comparison}

Previous mass-loss rates of IRC\,+10420 were derived from CO
\citep{CastroCarrizo2007,DinhVTrung2009} and from mid- and far-infrared imaging
\citep{Shenoy2016}, assuming radial expansion of all the circumstellar material.
They reach $\sim10^{-3}\,M_\odot$\,yr$^{-1}$ during high mass-loss episodes. Our
envelope rate is lower, because in our model most of the dust resides in the
equatorial components, for which we do not derive a rate. Expressed as a
mass-loss rate, the dust of these components would correspond to the high-rate
episodes inferred in those studies. The two pictures therefore differ mainly in
how the equatorial material is interpreted.

Radio observations show a molecular envelope that is approximately spherical on
large scales, whereas optical and infrared images show a structured, equatorially
concentrated nebula \citep{DinhVTrung2017}. Our model reconciles the two: a
spherical, expanding envelope that dominates at large radii, and equatorial
components that dominate the scattered light within $\approx3''$. The complex
inner ejecta in the HST images \citep{Humphreys1997,Tiffany2010} coincide with
the disk and the ring of the model, where the surface density of the scattering
grains is highest. The roughly spherical outer shells at $5''$--$6''$ lie beyond
the range of our $K'$ profiles and contribute little to the SED.

\subsection{Geometry of the ejecta} \label{sec:geometry}

In the meridional plane, the flared inner disk and the detached ring form a
bow-tie-shaped structure (Figure~\ref{fig:rz}a). The surface of the ring rises to
$\approx20$--$30^\circ$ above the equatorial plane. This leaves biconical,
hourglass-shaped regions around the poles that contain only the tenuous envelope.
Along the polar axis, the dust is optically thin and warmer than in the shielded
mid-plane (Figure~\ref{fig:rz} b, d). Most of the starlight that reaches the
observer passes through this polar region, which explains the moderate
circumstellar extinction ($A_V^{\rm cs}\approx1.85$\,mag;
Section~\ref{sec:extinction}) despite the large equatorial optical depth.

This geometry is consistent with the pole-on view inferred from HST spectroscopy
and proper motions \citep{Humphreys2002,Tiffany2010} and from integral-field
spectroscopy \citep{Davies2007}. A similar hourglass geometry, seen nearly
pole-on, was proposed for the Br$\gamma$ emission on au scales
\citep{Oudmaijer2013}. If so, the same axisymmetry would persist from
$\sim10$\,au to $\sim10^4$\,au, i.e., over the last $\sim10^3$\,yr of mass loss.
In the pole-on view, the circular isophotes and the centrosymmetric polarization
vectors are reproduced without any azimuthal structure. Viewed pole-on, a flared
equatorial structure and the walls of a bipolar cavity both appear as rings in
the images. The SED alone does not distinguish between them, but the polarimetry
requires the scattering surface to lie at low latitudes
($\lesssim20^\circ$--$30^\circ$), which excludes a narrow bipolar cavity; a
wide-angle cavity is nearly the same structure as the flared ring
(Section~\ref{sec:approach}).

\textit{Grain sizes}---The grains in the equatorial components (0.3--0.5\,$\mu$m)
are larger than those in the envelope ($\approx0.15\,\mu$m). Grains of a few
tenths of a micron have also been inferred from scattered-light polarimetry
around other evolved stars, including AGB stars and the red hypergiant VY~CMa
\citep[e.g.,][]{Norris2012,Scicluna2015}. In IRC\,+10420, the largest submicron
grains are found at the outer edge of the ring. If the ring consists of material
from the high mass-loss phase (Section~\ref{sec:massloss}), this material is the
oldest in the ring, and the larger grains may reflect a longer time for grain
growth; alternatively, the condensation conditions may have been different, for
example a higher density or a lower stellar temperature.

\subsection{Luminosity and evolutionary state} \label{sec:luminosity}

The anisotropy of the emission has a direct consequence for the location of
IRC\,+10420 in the Hertzsprung--Russell diagram. Previous luminosities were
derived from the integrated SED under the assumption of isotropic emission
\citep[e.g.,][]{Jones1993,Oudmaijer2013,Koumpia2022}. For the pole-on geometry
favored here, this overestimates the luminosity by a factor of $\approx1.5$
(Section~\ref{sec:sed}). At the \textit{Gaia} distance, our model gives
$L_\star\approx3.3\times10^5\,L_\odot$ ($\log L_\star/L_\odot\approx5.5$) instead
of $\approx5\times10^5\,L_\odot$. IRC\,+10420 would then lie clearly below the
empirical upper luminosity limit for cool supergiants, $\log L/L_\odot\approx5.8$
\citep{Humphreys1979,Humphreys1994}. This conclusion depends on the axisymmetric,
equatorially concentrated geometry and on the pole-on view. It also depends on
the division of the extinction (Section~\ref{sec:sed}): attributing all the
selective extinction to the interstellar medium would raise the luminosity by a
factor of 1.3.

A similar result was obtained for WOH G64 in the Large Magellanic Cloud: modeling
its SED and mid-infrared interferometry with an optically thick dusty torus
viewed close to pole-on, \citet{Ohnaka2008} derived a luminosity about two times
lower than previous estimates based on spherical shells, which brought the star
into agreement with the evolutionary tracks of a $\sim25\,M_\odot$ star. WOH G64
was recently reported to have become a yellow hypergiant
\citep{MunozSanchez2026}, but new spectra again show the TiO bands of a red
supergiant \citep{vanLoon2026}. In either case, it shares with IRC\,+10420 a
dusty, equatorially concentrated envelope seen nearly pole-on.

\subsection{Limitations of the model} \label{sec:limitations}

The adopted model is one of possibly many that reproduce the data
(Section~\ref{sec:approach}). Several simplifications of the model should be kept
in mind.
\begin{enumerate}[leftmargin=1em]
\setlength{\itemsep}{1pt}
\item The grains are spherical, compact silicate grains with a single set of
    optical constants. Non-spherical or porous grains, or other dust species
    such as alumina or metallic iron, would change the scattering matrices,
    and hence the grain sizes inferred from the polarization.
\item The density distribution is axisymmetric and smooth. It represents an
    azimuthal average of the clumpy, arc-like ejecta seen with HST.
\item The inclination is fixed at $i=0^\circ$. A nearly pole-on view is,
    however, well constrained by observations: HST spectroscopy and proper
    motions \citep{Humphreys2002,Tiffany2010}, integral-field spectroscopy
    \citep{Davies2007}, and the circular isophotes and centrosymmetric
    polarization of the $K'$ nebula. For $i\lesssim30^\circ$, the apparent
    luminosity changes by less than 5\% (Table~\ref{tab:fbol}), and the SED fit
    deteriorates at larger inclinations (Section~\ref{sec:sed}).
\item The star is represented by a blackbody (Section~\ref{sec:star}). Kurucz
    model atmospheres of the same temperature fit the optical photometry less
    well, probably because the continuum forms in the wind.
\item The disk of 300\,$\mu$m grains is required to reproduce the
    (sub)millimeter fluxes. Its mass, however, is uncertain by at least the
    uncertainty of the millimeter opacity of large grains. A contribution of
    free-free emission from the ionized wind at 1.3\,mm, which is not included
    in the model, would reduce the required mass.
\item The envelope mass-loss rate assumes a single, constant expansion velocity
    and a constant gas-to-dust ratio. The kinematics of the equatorial
    components are not constrained by the present data.
\end{enumerate}

\section{Summary} \label{sec:summary}

We have modeled the dusty circumstellar envelope of the yellow hypergiant
IRC\,+10420 with the three-dimensional Monte Carlo radiative transfer code
RADMC-3D, including full polarized scattering by spherical silicate grains. The
models were constrained simultaneously by the SED from $0.4\,\mu$m to 1.3\,mm and
by the $K'$ ($2.2\,\mu$m) imaging polarimetry of \citet{Shenoy2015}. Our main
results are as follows.

\begin{enumerate}[leftmargin=1em]
\setlength{\itemsep}{1pt}
\item The SED and the polarimetry are reproduced by a pole-on model with five
    dust components: a flared inner disk of $0.3\,\mu$m grains (300--3000\,au);
    a detached ring of $0.3\,\mu$m grains (3000--10\,000\,au), whose surface
    density rises as $r^{1.5}$ to a maximum at $10^4$\,au ($2''$); the outer
    edge of the ring (10\,000--16\,000\,au), which consists of $0.5\,\mu$m
    grains; a geometrically thin disk of $300\,\mu$m grains; and a spherical
    envelope of $0.15\,\mu$m grains with $\rho\propto r^{-2.3}$ ($\tau_V=4.9$).
    The central star has $L_\star=4.5\times10^5\,L_\odot$ and $T_\star=7000$\,K
    at $d=5$\,kpc, and the total dust mass is $7.4\times10^{-2}\,M_\odot$. We
    regard this multi-component structure as an azimuthally averaged
    representation of a structured equatorial outflow with a quasi-stationary,
    settled mid-plane, rather than as physically distinct disks.

\item The model fits the dereddened SED with $\sigma_{\log}=0.068$\,dex and
    $\chi^2_\nu=1.20$. The residuals are within 0.06\,dex from $B$ to $I$,
    0.13\,dex from 3 to $30\,\mu$m, and 0.19\,dex in the far-infrared and
    submillimeter. Stellar temperatures of 7000--8500\,K, which include the
    spectroscopic values, fit the SED equally well.

\item The model reproduces the observed radial profiles of the $K'$ total
    intensity, polarized intensity, and fractional polarization with rms
    residuals of 0.04--0.08\,dex. The fractional polarization peaks at 0.15 at
    $1\farcs7$ (observed: 0.14), and the polarization vectors are
    centrosymmetric. Scattered light contributes 44\%--74\% of the $K'$
    intensity between $0\farcs5$ and $2\farcs3$.

\item The polarimetry constrains the grain size in each component. The flat
    polarized-intensity profile between $0\farcs7$ and $1\farcs7$ arises from
    the rising surface density of the ring, and the decline beyond $2''$
    requires weakly polarizing $0.5\,\mu$m grains at the outer edge of the
    ring, beginning close to $10^4$\,au. Grains with $a_0\gtrsim0.7\,\mu$m
    would produce radially oriented polarization vectors and are excluded. The
    circumstellar color excess is set by the envelope grains and requires
    $a_0\approx0.15\,\mu$m for $T_\star=7000$\,K and 0.10--0.15\,$\mu$m for
    $T_\star=8000$--8500\,K; grains of $0.20\,\mu$m give poorer fits at all
    temperatures.

\item The total extinction toward the star is $A_V\approx6.9$\,mag, of which
    5\,mag is interstellar, consistent with diffuse interstellar band
    measurements, and $\approx1.9$\,mag is circumstellar. If all the selective
    extinction is attributed to the interstellar medium, an almost equally good
    fit requires $L_\star=6.0\times10^5\,L_\odot$.

\item The equatorial components absorb the starlight emitted toward low
    latitudes and re-emit and scatter it toward the poles, so that the model
    seen pole-on is 1.55 times brighter than an isotropic emitter. Luminosities
    derived from the integrated SED therefore overestimate $L_\star$ by a
    factor of $\approx1.5$. The model gives $L_\star=4.5\times10^5\,L_\odot$ at
    5\,kpc and $\approx3.3\times10^5\,L_\odot$ ($\log
    L_\star/L_\odot\approx5.5$) at the \textit{Gaia} distance of 4.26\,kpc,
    independent of $T_\star$ between 7000 and 8500\,K. This places IRC\,+10420
    well below the empirical upper luminosity limit for cool supergiants.

\item For $v_{\rm exp}=35$\,km\,s$^{-1}$, the envelope corresponds to a
    mass-loss rate of
    $\approx10^{-5}$--$10^{-4}\,(\psi/100)\,M_\odot$\,yr$^{-1}$. The equatorial
    components contain $6.3\times10^{-2}\,M_\odot$ of dust, about six times the
    envelope mass; because their kinematics are unknown, we do not derive a
    mass-loss history from them. If the wind velocity is taken as an upper
    limit to the outward velocity of the equatorial material, the ring is at
    least $\approx1350$--2200\,yr old (for $d=5$\,kpc) and was formed at an
    equatorial rate of at most
    $\approx2\times10^{-3}\,(\psi/100)\,M_\odot$\,yr$^{-1}$, consistent with
    the high mass-loss phase that ended $\sim2000$\,yr ago \citep{Shenoy2016}.

\item In the meridional plane, the flared inner disk and the ring form a
    bow-tie structure with biconical, low-density regions around the poles.
    This resembles, on scales $\sim10^3$ times larger, the hourglass geometry
    proposed for the Br$\gamma$ emission on au scales \citep{Oudmaijer2013}.
    Viewed pole-on, a flared equatorial structure and the walls of a bipolar
    cavity both appear as rings; the SED does not distinguish between them, but
    the polarimetry requires the scattering surface to lie at low latitudes,
    which excludes a narrow bipolar cavity; a wide-angle cavity is nearly the
    same structure as the flared ring. Near-infrared imaging polarimetry at
    higher angular resolution, e.g., with extreme adaptive optics, would test
    the predicted polarization inside $0\farcs4$, where the MMT-Pol data are
    saturated.

\item Subtracting a stellar PSF scaled at $0\farcs5$, as in the observational
    analysis, removes most of the nebular light close to the star. Applied to
    the model, it gives $I^{\rm neb}/I^{\rm raw}=0.01$ at $0\farcs7$, whereas
    the true fraction is 0.44. Nebular intensities and polarizations derived in
    this way within $\approx1''$ should therefore be interpreted with caution.
\end{enumerate}

\section*{Acknowledgments}

I am deeply grateful to Terry Jones and Roberta Humphreys for inspiring this work
and for their helpful comments. I thank the anonymous reviewer for constructive
comments and suggestions. This research was supported by Basic Science Research
Program through the National Research Foundation of Korea (NRF) funded by the
Ministry of Education (RS-2022-NR075638). This research has made use of the
SIMBAD database and VizieR catalogue access tool, operated at CDS, Strasbourg,
France.

\appendix

\section{Long-term optical photometry}\label{sec:ltphot}
\restartappendixnumbering

\begin{figure}
\smallplot{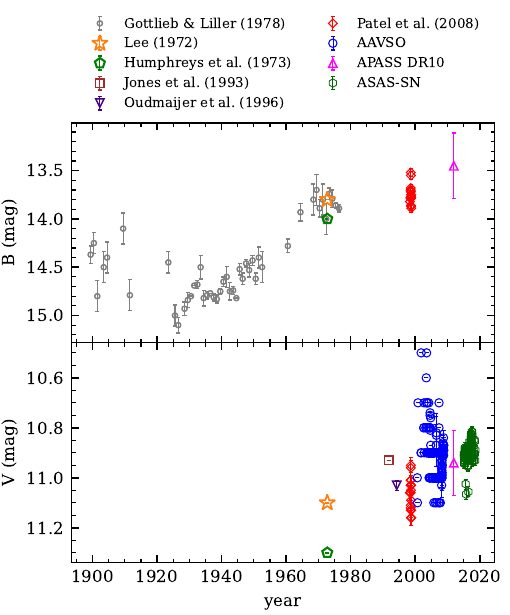}
\caption{\textbf{Long-term $B$ and $V$ photometry of IRC\,+10420
(Table~\ref{tab:ltphot}). Error bars are the published uncertainties, the error
bars plotted by \citet{Gottlieb1978} for the Harvard annual means, and the scatter
of 10 nights for APASS DR10. No uncertainties are available for the measurements
of 1972 and 1991 and for most of the AAVSO measurements, which are shown without
error bars. The ASAS-SN measurements have uncertainties of 0.02\,mag.}
\label{fig:ltphot}}
\end{figure}

Table~\ref{tab:ltphot} and Figure~\ref{fig:ltphot} collect the published $B$ and
$V$ photometry of IRC\,+10420 since its discovery, together with the survey data
of the AAVSO International Database, APASS DR10, and ASAS-SN. Annual means from
283 Harvard photographic plates (digitized from Figure~3 of
\citealt{Gottlieb1978}) show that the star varied irregularly between $B=14$ and
15 before 1925 and then brightened at $\approx0.025$\,mag\,yr$^{-1}$, with a
possible slowing of the brightening in the late 1970s \citep{Gottlieb1978}; see
also \citet{Jones1993}. Since 1972, in contrast, the optical brightness has
remained nearly constant, at $V\approx10.9$--11.3 and $B\approx13.4$--14.0
\citep{Humphreys1973,Jones1993,Oudmaijer1996,Patel2008}. CCD $V$ measurements in
the AAVSO International Database fill the gap between 1998 and 2012 and give
$V=10.92$ with an rms scatter of 0.10\,mag for 2000--2008. The ASAS-SN $V$-band
photometry of 2015--2018 (128 epochs) gives a mean of $V=10.89$ with an rms
scatter of 0.04\,mag, and the star is classified as an irregular variable
\citep{Shappee2014,Jayasinghe2019}, although it may be saturated in these data.
The scatter within a single season is up to 0.2\,mag in $V$ and 0.4\,mag in $B$
\citep{Patel2008}, and the measurements were obtained with different telescopes
and photometric systems, so the differences of 0.1--0.3\,mag between epochs are
not significant. The near-infrared fluxes, in contrast, decreased by
$\approx0.75$\,mag between 1975 and 1983 \citep{Jones1993,Oudmaijer1996}. The
constant optical brightness since the 1970s is consistent with the nearly
constant spectral type since 1994 \citep{Koumpia2022} and supports the use of a
single stellar model for the SED data of 1983--2007 (Section~\ref{sec:obs_sed}).

\begin{deluxetable}{llcccc}
\tablecaption{Long-term Optical Photometry of IRC\,+10420}\label{tab:ltphot}
\tablehead{\colhead{Source} & \colhead{Band} & \colhead{Epoch} & \colhead{$N$} & \colhead{Mean} & \colhead{Range} \\
\colhead{} & \colhead{} & \colhead{(yr)} & \colhead{} & \colhead{(mag)} & \colhead{(mag)}}
\startdata
\citet{Gottlieb1978}\tablenotemark{a} & $B$ & 1899--1976 & 46 & 14.46 & 13.70--15.10 \\
Lee (1972)\tablenotemark{b} & $B$ & 1972 & 1 & 13.80 & \nodata \\
Lee (1972)\tablenotemark{b} & $V$ & 1972 & 1 & 11.10 & \nodata \\
\citet{Humphreys1973} & $B$ & 1972 & 1 & 14.00 & \nodata \\
\citet{Humphreys1973} & $V$ & 1972 & 1 & 11.30 & \nodata \\
\citet{Jones1993} & $V$ & 1991 & 1 & 10.93 & \nodata \\
\citet{Jones1993} & $R$ & 1991 & 1 & 9.56 & \nodata \\
\citet{Jones1993} & $I$ & 1991 & 1 & 7.95 & \nodata \\
\citet{Oudmaijer1996} & $V$ & 1994 & 1 & 11.03 & \nodata \\
\citet{Patel2008} & $B$ & 1998 & 18 & 13.75 & 13.52--13.89 \\
\citet{Patel2008} & $V$ & 1998 & 18 & 11.07 & 10.95--11.16 \\
AAVSO\tablenotemark{c} & $V$ & 2000--2008 & 203 & 10.92 & 10.50--11.10 \\
APASS DR10 \citep{Henden2018} & $B$ & 2012 & 1 & 13.45 & \nodata \\
APASS DR10 \citep{Henden2018} & $V$ & 2012 & 1 & 10.94 & \nodata \\
ASAS-SN\tablenotemark{d} & $V$ & 2015--2018 & 128 & 10.89 & 10.81--11.07 \\
\enddata
\tablecomments{Apparent magnitudes (not corrected for extinction). Individual measurements are shown in Figure~\ref{fig:ltphot}. APASS DR10 values are means of 10 nights.}
\tablenotetext{a}{Annual means of 283 Harvard photographic plates, digitized from Figure 3 of \citet{Gottlieb1978}; $N$ is the number of annual means.}
\tablenotetext{b}{Private communication, as listed by \citet{Jones1993}.}
\tablenotetext{c}{AAVSO International Database, CCD $V$ measurements of three observers (not transformed); ten measurements of a fourth observer, with a scatter of 0.5\,mag, and one discrepant measurement ($V=12.0$) are not used.}
\tablenotetext{d}{ASAS-SN catalog of variable stars \citep{Shappee2014,Jayasinghe2019}: ASASSN-V\,J192648.08+112118.0, classified as an irregular variable (L); the star may be saturated in the ASAS-SN $V$-band data.}
\end{deluxetable}

\begin{thebibliography}{}

\bibitem[{Bailer-Jones} et al.(2021)]{BailerJones2021} Bailer-Jones, C. A. L.,
    Rybizki, J., Fouesneau, M., Demleitner, M., \& Andrae, R. 2021, \aj, 161, 147

\bibitem[{Beichman} et al.(1988)]{Beichman1988} Beichman, C. A., Neugebauer, G.,
    Habing, H. J., Clegg, P. E., \& Chester, T. J. 1988, Infrared Astronomical
    Satellite (IRAS) Catalogs and Atlases, Vol.~1: Explanatory Supplement, NASA
    RP-1190 (Washington, DC: NASA)

\bibitem[{Bl{\"o}cker} et al.(1999)]{Blocker1999} Bl{\"o}cker, T., Balega, Y.,
    Hofmann, K.-H., et al. 1999, \aap, 348, 805

\bibitem[{Bohren} \& {Huffman}(1983)]{Bohren1983} Bohren, C. F., \& Huffman, D.
    R. 1983, Absorption and Scattering of Light by Small Particles (New York:
    Wiley)

\bibitem[{Bujarrabal} et al.(2013)]{Bujarrabal2013} Bujarrabal, V., Alcolea, J.,
    Van Winckel, H., Santander-Garc{\'i}a, M., \& Castro-Carrizo, A. 2013, \aap,
    557, A104

\bibitem[{Cardelli} et al.(1989)]{Cardelli1989} Cardelli, J. A., Clayton, G. C.,
    \& Mathis, J. S. 1989, \apj, 345, 245

\bibitem[{Castro-Carrizo} et al.(2007)]{CastroCarrizo2007} Castro-Carrizo, A.,
    Quintana-Lacaci, G., Bujarrabal, V., Neri, R., \& Alcolea, J. 2007, \aap,
    465, 457

\bibitem[{Davies} et al.(2007)]{Davies2007} Davies, B., Oudmaijer, R. D., \&
    Sahu, K. C. 2007, \apj, 671, 2059

\bibitem[{de Jager}(1998)]{deJager1998} de Jager, C. 1998, \aapr, 8, 145

\bibitem[{de Ruyter} et al.(2006)]{deRuyter2006} de Ruyter, S., Van Winckel, H.,
    Maas, T., et al. 2006, \aap, 448, 641

\bibitem[{de Wit} et al.(2008)]{deWit2008} de Wit, W. J., Oudmaijer, R. D.,
    Fujiyoshi, T., et al. 2008, \apjl, 685, L75

\bibitem[{Di Francesco} et al.(2008)]{DiFrancesco2008} Di Francesco, J.,
    Johnstone, D., Kirk, H., et al. 2008, \apjs, 175, 277

\bibitem[{Dinh-V-Trung} et al.(2009)]{DinhVTrung2009} Dinh-V-Trung, Muller, S.,
    Lim, J., Kwok, S., \& Muthu, C. 2009, \apj, 697, 409

\bibitem[{Dinh-V-Trung} et al.(2017)]{DinhVTrung2017} Dinh-V-Trung, Wong, K. T.,
    \& Lim, J. 2017, \apj, 849, 44

\bibitem[{Dullemond} et al.(2012)]{Dullemond2012} Dullemond, C. P., Juhasz, A.,
    Pohl, A., et al. 2012, RADMC-3D: A Multi-purpose Radiative Transfer Tool,
    Astrophysics Source Code Library, ascl:1202.015

\bibitem[{Egan} et al.(2003)]{Egan2003} Egan, M. P., Price, S. D., Kraemer, K.
    E., et al. 2003, The Midcourse Space Experiment Point Source Catalog Version
    2.3, AFRL-VS-TR-2003-1589 (Springfield, VA: NTIS)

\bibitem[{Giguere} et al.(1976)]{Giguere1976} Giguere, P. T., Woolf, N. J., \&
    Webber, J. C. 1976, \apjl, 207, L195

\bibitem[{Gottlieb} \& {Liller}(1978)]{Gottlieb1978} Gottlieb, E. W., \&
    Liller, W. 1978, \apj, 225, 488

\bibitem[{Henden} et al.(2016)]{Henden2016} Henden, A. A., Templeton, M.,
    Terrell, D., et al. 2016, VizieR Online Data Catalog, II/336

\bibitem[{Henden} et al.(2018)]{Henden2018} Henden, A. A., Levine, S.,
    Terrell, D., et al. 2018, AAS Meeting Abstracts, 232, 223.06

\bibitem[{H{\"o}fner}(2008)]{Hofner2008} H{\"o}fner, S. 2008, A\&A, 491, L1

\bibitem[{Humphreys} \& {Davidson}(1979)]{Humphreys1979} Humphreys, R. M., \&
    Davidson, K. 1979, \apj, 232, 409

\bibitem[{Humphreys} \& {Davidson}(1994)]{Humphreys1994} Humphreys, R. M., \&
    Davidson, K. 1994, \pasp, 106, 1025

\bibitem[{Humphreys} et al.(2002)]{Humphreys2002} Humphreys, R. M., Davidson, K.,
    \& Smith, N. 2002, \aj, 124, 1026

\bibitem[{Humphreys} et al.(1973)]{Humphreys1973} Humphreys, R. M., Strecker, D.
    W., Murdock, T. L., \& Low, F. J. 1973, \apjl, 179, L49

\bibitem[{Humphreys} et al.(1997)]{Humphreys1997} Humphreys, R. M., Smith, N.,
    Davidson, K., et al. 1997, \aj, 114, 2778

\bibitem[{Humphreys} et al.(2007)]{Humphreys2007} Humphreys, R. M., Helton, L.
    A., \& Jones, T. J. 2007, \aj, 133, 2716

\bibitem[{Indebetouw} et al.(2005)]{Indebetouw2005} Indebetouw, R., Mathis, J.
    S., Babler, B. L., et al. 2005, \apj, 619, 931

\bibitem[{Jayasinghe} et al.(2019)]{Jayasinghe2019} Jayasinghe, T., Stanek,
    K. Z., Kochanek, C. S., et al. 2019, \mnras, 486, 1907

\bibitem[{Jones} et al.(1993)]{Jones1993} Jones, T. J., Humphreys, R. M., Gehrz,
    R. D., et al. 1993, \apj, 411, 323

\bibitem[{Kastner} \& {Weintraub}(1995)]{KastnerWeintraub1995} Kastner, J. H., \&
    Weintraub, D. A. 1995, \apj, 452, 833

\bibitem[{Kawada} et al.(2007)]{Kawada2007} Kawada, M., Baba, H., Barthel, P. D.,
    et al. 2007, \pasj, 59, S389

\bibitem[{Kenyon} \& {Hartmann}(1987)]{Kenyon1987} Kenyon, S. J., \& Hartmann, L.
    1987, \apj, 323, 714

\bibitem[{Klochkova} et al.(1997)]{Klochkova1997} Klochkova, V. G., Chentsov, E.
    L., \& Panchuk, V. E. 1997, \mnras, 292, 19

\bibitem[{Klochkova} et al.(2002)]{Klochkova2002} Klochkova, V. G., Yushkin, M.
    V., Chentsov, E. L., \& Panchuk, V. E. 2002, Astronomy Reports, 46, 139

\bibitem[{Klochkova} et al.(2016)]{Klochkova2016} Klochkova, V. G., Chentsov, E.
    L., Miroshnichenko, A. S., Panchuk, V. E., \& Yushkin, M. V. 2016, \mnras,
    459, 4183

\bibitem[{Koumpia} et al.(2022)]{Koumpia2022} Koumpia, E., Oudmaijer, R. D., de
    Wit, W. J., et al. 2022, \mnras, 515, 2766

\bibitem[{Kurucz}(1992)]{Kurucz1992} Kurucz, R. L. 1992, in IAU Symp. 149, The
    Stellar Populations of Galaxies, ed. B. Barbuy \& A. Renzini (Dordrecht:
    Kluwer), 225

\bibitem[{Lloyd Evans}(1990)]{LloydEvans1990} Lloyd Evans, T. 1990, \mnras, 243,
    336

\bibitem[{Mathis} et al.(1977)]{Mathis1977} Mathis, J. S., Rumpl, W., \&
    Nordsieck, K. H. 1977, ApJ, 217, 425

\bibitem[{Mauron} \& {Josselin}(2011)]{Mauron2011} Mauron, N., \& Josselin, E.
    2011, A\&A, 526, A156

\bibitem[{Moffat}(1969)]{Moffat1969} Moffat, A. F. J. 1969, \aap, 3, 455

\bibitem[{Molster} et al.(2001)]{Molster2001} Molster, F. J., Yamamura, I.,
    Waters, L. B. F. M., et al. 2001, \aap, 366, 923

\bibitem[{Morris}(1987)]{Morris1987} Morris, M. 1987, \pasp, 99, 1115

\bibitem[{Mu{\~n}oz-Sanchez} et al.(2026)]{MunozSanchez2026} Mu{\~n}oz-Sanchez,
    G., Kalitsounaki, M., de Wit, S., et al. 2026, NatAs, 10, 702

\bibitem[{Nieuwenhuijzen} \& {de Jager}(2000)]{Nieuwenhuijzen2000}
    Nieuwenhuijzen, H., \& de Jager, C. 2000, \aap, 353, 163

\bibitem[{Norris} et al.(2012)]{Norris2012} Norris, B. R. M., Tuthill, P. G.,
    Ireland, M. J., et al. 2012, \nat, 484, 220

\bibitem[{Ohnaka} et al.(2008)]{Ohnaka2008} Ohnaka, K., Driebe, T., Hofmann,
    K.-H., Weigelt, G., \& Wittkowski, M. 2008, A\&A, 484, 371

\bibitem[{Oudmaijer}(1998)]{Oudmaijer1998} Oudmaijer, R. D. 1998, \aaps, 129, 541

\bibitem[{Oudmaijer} \& {de Wit}(2013)]{Oudmaijer2013} Oudmaijer, R. D., \& de
    Wit, W. J. 2013, \aap, 551, A69

\bibitem[{Oudmaijer} et al.(1996)]{Oudmaijer1996} Oudmaijer, R. D., Groenewegen,
    M. A. T., Matthews, H. E., Blommaert, J. A. D. L., \& Sahu, K. C. 1996,
    \mnras, 280, 1062

\bibitem[{Patel} et al.(2008)]{Patel2008} Patel, M., Oudmaijer, R. D., Vink, J.
    S., et al. 2008, \mnras, 385, 967

\bibitem[{Scicluna} et al.(2015)]{Scicluna2015} Scicluna, P., Siebenmorgen, R.,
    Wesson, R., et al. 2015, \aap, 584, L10

\bibitem[{Shappee} et al.(2014)]{Shappee2014} Shappee, B. J., Prieto, J. L.,
    Grupe, D., et al. 2014, \apj, 788, 48

\bibitem[{Shenoy} et al.(2015)]{Shenoy2015} Shenoy, D. P., Jones, T. J., Packham,
    C., \& Lopez-Rodriguez, E. 2015, \aj, 150, 15

\bibitem[{Shenoy} et al.(2016)]{Shenoy2016} Shenoy, D., Humphreys, R. M., Jones,
    T. J., et al. 2016, \aj, 151, 51

\bibitem[{Skrutskie} et al.(2006)]{Skrutskie2006} Skrutskie, M. F., Cutri, R. M.,
    Stiening, R., et al. 2006, \aj, 131, 1163

\bibitem[{Sloan} et al.(2003)]{Sloan2003} Sloan, G. C., Kraemer, K. E., Price, S.
    D., \& Shipman, R. F. 2003, \apjs, 147, 379

\bibitem[{Smith}(2014)]{Smith2014} Smith, N. 2014, \araa, 52, 487

\bibitem[{Suh}(1999)]{Suh1999} Suh, K.-W. 1999, \mnras, 304, 389

\bibitem[{Suh}(2016)]{Suh2016} Suh, K.-W. 2016, \apj, 819, 61

\bibitem[{Tiffany} et al.(2010)]{Tiffany2010} Tiffany, C., Humphreys, R. M.,
    Jones, T. J., \& Davidson, K. 2010, \aj, 140, 339

\bibitem[{Trujillo} et al.(2001)]{Trujillo2001} Trujillo, I., Aguerri, J. A. L.,
    Cepa, J., \& Guti{\'e}rrez, C. M. 2001, \mnras, 328, 977

\bibitem[{van Loon} \& {Ohnaka}(2026)]{vanLoon2026} van Loon, J. Th., \&
    Ohnaka, K. 2026, \mnras, 546, 1

\bibitem[{Walmsley} et al.(1991)]{Walmsley1991} Walmsley, C. M., Chini, R.,
    Kreysa, E., et al. 1991, \aap, 248, 555

\bibitem[{Yamamura} et al.(2000)]{Yamamura2000} Yamamura, I., Dominik, C., de
    Jong, T., Waters, L. B. F. M., \& Molster, F. J. 2000, \aap, 363, 629

\bibitem[{Yamamura} et al.(2010)]{Yamamura2010} Yamamura, I., Makiuti, S., Ikeda,
    N., et al. 2010, VizieR Online Data Catalog, II/298

\end{thebibliography}
\end{document}